\documentclass[a4paper,12pt]{article}

\usepackage{graphicx,amssymb,float,placeins}
\usepackage{mathrsfs}
\usepackage{amsmath,fancyhdr,lastpage}
\usepackage{tikz}
\usepackage{soul,color}
\usepackage[normalem]{ulem}
\usepackage{subfigure}

\newcommand{\mb}{\mathbf}

\usetikzlibrary{decorations.markings}
\usetikzlibrary{decorations.pathmorphing}
\usetikzlibrary{fit}
\usetikzlibrary{calc,arrows,patterns}
\usepgflibrary{plotmarks}

\usepackage{paralist}

\newcommand{\tr}{\mathrm{tr}\,}

\newcommand{\be}{\begin{equation}}
\newcommand{\en}{\end{equation}}
\newcommand{\wnh}{W_{\rm nH}}
\newcommand{\im}{I_{\rm m}}
\newcommand{\tim}{\frac{3}{\im}}
\newcommand{\iim}{\frac{I_1}{\im}}
\newcommand{\wgent}{W_{\rm Gent}}
\newcommand{\wcohen}{W_{\rm Cohen}}
\newcommand{\wnew}{W_{\rm New}}
\newcommand{\wbeatty}{W_{\rm Beatty}}
\newcommand{\betah}{\hat{\beta}}
\newcommand{\invL}{\mathscr{L}^{-1}(x)}

\newcommand{\wJG}{W_{\rm 3ch}}

\newcommand{\wab}{W_{\rm 8ch}}

\newcommand{\rr}{{\rm r}}

\title{A comparison of limited-stretch  models of rubber elasticity }
\author{S. R. Rickaby\thanks{Email: stephen.r.rickaby@gmail.com}, 
N. H. Scott\thanks{Email: n.scott@uea.ac.uk}\\
School of Mathematics, University of East Anglia,\\
 Norwich Research Park, Norwich NR4 7TJ
}

\date{}

\begin{document}
\maketitle

\thispagestyle{fancy} \lhead{\emph{International Journal of Non-Linear Mechanics}   (2015) {\bf 68},  71--86.    \\   
doi:10.1016/j.ijnonlinmec.2014.06.009\\ 
Published online 23 June 2014\\
arXiv: 
}
\chead{14 May 2020}
\rhead{Page\  \thepage\ of\ \pageref{LastPage}}
\cfoot{}

\begin{abstract}
In this paper we describe various limited-stretch models of nonlinear rubber elasticity, each dependent on only the first invariant of the left Cauchy-Green strain tensor and having only two independent material constants.  The models are described as limited-stretch, or restricted elastic, because the  strain energy and  stress response   become infinite at a finite value of the first invariant.  These models describe well the limited stretch of the polymer chains of which rubber is composed.   We discuss  Gent's model which is the simplest limited-stretch model and agrees well with experiment.  Various statistical models are then described:   the one-chain, three-chain,  four-chain and Arruda-Boyce eight-chain models, all of which involve the inverse Langevin function.  A  numerical  comparison between the three-chain and eight-chain models is provided. Next, we compare various models which involve approximations to the inverse Langevin function with the exact inverse Langevin function of the eight-chain model.  A new approximate model is proposed that is as simple as Cohen's original model but significantly more accurate.  We show that effectively the eight-chain model may be regarded as a linear combination of the neo-Hookean and Gent models.  Treloar's model is shown to have about half the percentage error of our new model but it is much more complicated.  For completeness a modified Treloar model is introduced but this is only slightly more accurate than Treloar's original model.
For the deformations of uniaxial tension, biaxial tension, pure shear and simple shear we compare the accuracy of these models, and that of Puso, with the eight-chain model by means of graphs and a table.  Our approximations compare extremely well with models frequently used and described in the literature, having the smallest mean percentage error over most of the range of the argument. 

\vspace{1mm}\noindent
\textbf{Keywords}
Inverse Langevin function, Strain energy, Limited stretch, Restricted elastic, FJC and WLC models, Biological models\\
\textbf{MSC codes:}  74B20 $\cdot$ 74D10 $\cdot$ 74L15
\end{abstract}

\section{Introduction} % section 1
Several strain energy models of rubber elasticity are developed in this paper, some based on positing a form of the strain energy function, some based on statistical mechanical considerations of the polymer chains of which rubber is composed and others based on approximations to the statistical mechanics models.   The statistical mechanics models we consider in this paper involve only the first  invariant $I_1$ of the left Cauchy-Green strain tensor.   The simplest statistical model is based on Gaussian statistics and leads to the well known neo-Hookean strain energy function which depends linearly on $I_1$ and  on only one material constant, the shear modulus.  It is a feature of rubber elasticity that when a specimen of rubber is stretched in any direction a maximum stretch is reached, corresponding to the polymer chains being stretched to their maximum extent.  We model this property by requiring the stress response and strain energy to become infinite as this maximum stretch is reached.  Such a model of rubber elasticity is said to be a limited-stretch, or restricted elastic model.  All the other models we discuss are of this type and, furthermore, depend  on only two material constants, a shear modulus and the value $I_{\rm m}$ of the first invariant when maximum stretch is reached. Gent \cite{gent2005} discusses the relevance of modelling rubber using only the first principal invariant $I_1$ with its maximum $\im$.   Horgan and Saccomandi \cite{horgan2003} have shown that this approach of limiting chain extensibility may be used to model biological material.

All the models considered in this paper are freely jointed chain (FJC) models.  The alternative  worm-like chain (WLC) model is discussed briefly at the end of Section 3.

We consider the strain energy and stress response for each of the following models:   Gent \cite{gent}, Beatty \cite{beatty2008}, Van der Waals \cite{kilian} and Warner \cite{warner}.  None of these is directly related to statistical mechanics but each has limited stretch and depends on only two elastic constants.

More sophisticated statistical modelling also leads to limited-stretch models with two elastic constants, each depending on the inverse Langevin function  and deriving its limited-stretch behaviour from the singularity of this function.
Kuhn and Gr\"un \cite{kuhn}  used statistical mechanics to derive  an expression for the strain energy function of a single polymer chain which involved  the inverse Langevin function. A similar approach has been used to develop  network models based on cell structures,  including the James and Guth \cite{james} three-chain model, the Wang and Guth \cite{wang} four-chain model and Arruda and Boyce \cite{arruda} eight-chain model.   Wu and van der Giessen \cite{wu} presented a full network model.

\pagestyle{fancy}
\lhead{Models of rubber elasticity} \chead{Page\  \thepage\ of\ \pageref{LastPage}} \rhead{S. R. Rickaby, N. H. Scott}
\lfoot{} \cfoot{} \rfoot{}

When using any of the above  models an approximation to the inverse Langevin function is required.  
Perhaps the simplest approximations are those obtained by truncating the Taylor series.  However, many terms of the Taylor series are needed to approach convergence, see Itskov et al.\@ \cite{itskov} and Itskov et al.\@ \cite{itskov2011} for further discussion. Most approximations involve the Taylor series, such as the method of  Pad\'e approximants or further approximations to these.  Horgan and Saccomandi \cite{horgan2006} note that such   methods may be used to capture correctly the real singularities of the inverse Langevin function.    Cohen \cite{cohen} derived an approximation based on the $[3/2]$ Pad\'e approximant of the inverse Langevin function and  Treloar \cite{treloar1975} obtained a rational approximation to the inverse Langevin function which is related to the $[1/6]$ Pad\'e approximant of its Taylor series.  We present a new model, and a modified Treloar model, which are based on Pad\'e approximants of the reduced Langevin function, a function defined by multiplying out the simple poles of the inverse Langevin function.   Puso's \cite{puso} model does not appear to depend directly on  Pad\'e approximants. We include also a discussion on the additive removal of the real singularities of the inverse Langevin function.

Z\'{u}\~{n}iga and Beatty \cite{zuniga2002} and Beatty \cite{beatty2003} describe the James and Guth \cite{james} three-chain model, the Arruda and Boyce \cite{arruda} eight-chain model and the Wu and van der Giessen \cite{wu} full network model.  Beatty \cite{beatty2008} discuses the derivation of  the Cohen \cite{cohen}, Treloar \cite{treloar1975} and Horgan and Saccomandi \cite{horgan2002} approximations and concludes that the approximation of Treloar \cite{treloar1975} is the most accurate over the entire range of its argument.  Boyce \cite{boyce1996} directly compares the Gent and eight-chain models concluding that the eight-chain model gives a better interpretation of the physics of the polymer chain network, though both models provide excellent agreement with experimental data.

This paper is structured as follows. In Section \ref{sec:strainenergy} we define the Cauchy stress for an incompressible isotropic elastic material and specialize to the case where the strain energy depends only on the first  invariant, $I_1$, of the left Cauchy-Green strain tensor.  We derive expressions for the stress in the following four homogeneous deformations: uniaxial tension, biaxial tension, pure shear and simple shear. 
In Section \ref{sec:various} we discuss several models for limited-stretch rubber elasticity that are dependent on the first invariant only.   These models are: the neo-Hookean, Gent, Beatty, van der Waals and Warner models. The stress response and strain energy are given for each model.
 In Section \ref{sec:function} we define the Langevin function and its inverse $\mathscr{L}^{-1}(x)$ and present series expansions for them.  We introduce a reduced inverse Langevin function $f(x)$ which consists of the   inverse Langevin function $\mathscr{L}^{-1}(x)$ with its simple poles  removed by multiplying them out.  A series expansion for $f(x)$ is also given. 
 In Section \ref{sec:A-B} we describe several models of limited-stretch rubber elasticity which are based on the  inverse Langevin function, namely, the single-chain,  three-chain,    eight-chain and four-chain models. We note also in this Section that Beatty \cite{beatty2003} demonstrates an alternative derivation of the eight-chain model without reference to the eight-chain cell structure. We conclude this Section with a numerical comparison between the three-chain and eight-chain models.
 In Section \ref{sec:approcinvlan} we discuss several models which are based on approximations to the inverse Langevin function.  The first model consists of various truncations of the power series of $\mathscr{L}^{-1}(x)$.  This series and the series for $f(x)$ play an important role in the further models introduced, namely, those of Cohen, a new model, Treloar's model, a modification of Treloar's model, Puso's model, Indei et al.\@'s model and a model based on the additive removal of the real singularities of $\mathscr{L}^{-1}(x)$.   For most of these models we present the stress response and strain energy. We show that to a high degree of accuracy the eight-chain model may be regarded as a linear combination of the neo-Hookean and Gent models. 
 In Section \ref{sec:numerical} we give a numerical comparison of the various models, taking as reference the Arruda-Boyce \cite{arruda}  eight-chain model.  We first provide  a graphical  comparison of the neo-Hookean, Gent, Beatty, van der Waals and Warner models with the Arruda-Boyce eight-chain model, comparing the stress responses and strain energies.  We then proceed to compare graphically  Cohen's model, the new model, Treloar's model and the modified Treloar model with the Arruda-Boyce eight-chain model.  We consider the stress response, the strain energy and each of the four homogeneous deformations discussed in Section \ref{sec:strainenergy}.  We also give the mean percentage errors for all these quantities in a table.
 Finally, there is a discussion of the results  in Section~\ref{sec:conclusion}.

\section{Four homogeneous deformations} % section 2
\label{sec:strainenergy}

The Cauchy stress in an incompressible isotropic elastic material is given by
\be \label{eq1}
\mb{T} = -p\mb{I}+\beta \mb{B} +\beta_{-1}\mb{B}^{-1}
\en
where $p$ is an arbitrary pressure and $\mb{B}=\mb{F}\mb{F}^{\rm T}$ is the left Cauchy-Green strain tensor with $\mb{F}$ denoting the deformation gradient.  The response functions are given in terms of the strain energy $W$ by
\be \label{eq2} \beta  = 2\frac{\partial W}{\partial I_1},\quad \beta_{-1}  = -2\frac{\partial W}{\partial I_2} \en
where $I_1=\tr\mb{B}$ and $I_2=\tr\mb{B}^{-1}$ are the first two principal invariants of  $\mb{B}$.  Because of incompressibility the third principal invariant is given by $I_3=\det\mb{B}=1$.   We are assuming no dependence on  $I_2$  and so must  take $\beta_{-1}=0$ and $\beta=\beta(I_1)$.  Therefore, throughout this paper, the Cauchy stress (\ref{eq1}) reduces to
\be\label{eq3} \mb{T} = -p\mb{I}+\beta \mb{B}, \en
where the stress response $\beta$ is given by Eq.\@ (\ref{eq2})$_1$.

A method alternative to Eq.\@ (\ref{eq3})  of designating the stress in an incompressible isotropic elastic material is to take the strain energy to be a symmetric function of the principal stretches, $W=\widehat{W}(\lambda_1, \lambda_2, \lambda_3)$, and observe that the principal Cauchy stresses are given by
\be \label{eq4}  T_j = -p+\lambda_j\frac{\partial \widehat{W}}{\partial \lambda_j},\quad\mbox{for}\quad j=1, 2, 3,
\en
in which the principal stretches are denoted by $\lambda_j$, for $j=1,2,3$.  In terms of the principal stretches we have
\be \label{zz5} I_1=\tr\mb{B} = \lambda_1^2+\lambda_2^2+\lambda_3^2.   \en

We now examine four different homogeneous deformations and in each case compute the stress in terms of the largest principal stretch $\lambda>1$ and the response function $\beta$.

\subsection{Uniaxial tension}
We consider a uniaxial tension in the 1-direction with corresponding principal stretch $\lambda_1=\lambda>1$, so that incompressibility forces the other two principal stretches  to be $\lambda_2=\lambda_3=\lambda^{-1/2}$.  We take $p=\lambda^{-1}\beta$ to ensure the vanishing of the lateral stresses and then the only non-zero stress is the uniaxial tension
\be\label{eq5} T_{11}^{\rm uni}(\lambda)=(\lambda^2-\lambda^{-1})\beta,\quad I_1=\lambda^2+2\lambda^{-1}. \en

\subsection{Biaxial tension}
We consider a biaxial tension with equal principal stretches $\lambda_1=\lambda_2=\lambda>1$ in each of the 1- and 2-directions.  Incompressibility  forces the third principal stretch to be $\lambda_3=\lambda^{-2}$.  We take $p=\lambda^{-4}\beta$ to ensure that $T_{33}=0$ and then the only two non-zero stresses are the (equal) biaxial tensions in each of the 1- and 2-directions,  given by
\be\label{eq6} T_{11}^{\rm bi}(\lambda)=(\lambda^2-\lambda^{-4})\beta,\quad I_1=2\lambda^2+\lambda^{-4}. \en

\subsection{Pure shear}
We consider the pure shear deformation $\lambda_1=\lambda>1, \lambda_2=1, \lambda_3=\lambda^{-1}$ and take 
$p=\lambda^{-2}\beta $  so that $T_{33}=0$.  Then the largest stress is
\be\label{eq7} T_{11}^{\rm ps}(\lambda)=(\lambda^2-\lambda^{-2})\beta,
\quad I_1=\lambda^2+1+\lambda^{-2}. \en 
The only other non-zero stress is $T_{22}^{\rm ps}(\lambda)=(1-\lambda^{-2})\beta$. If $\lambda>1$ both these stresses are positive but if $\lambda<1$ both are negative, though $T_{11}^{\rm ps}(\lambda)$ remains the greater in absolute value since $T_{11}^{\rm ps}(\lambda)=(1+\lambda^2)T_{22}^{\rm ps}(\lambda)$.

\subsection{Simple shear}
We consider the simple shear deformation with deformation gradient
\[ \mb{F}=\begin{pmatrix}1&\gamma&0\\0&1&0\\0&0&1\end{pmatrix},
\mbox{\quad which implies that\quad}
 \mb{B}=\begin{pmatrix}1+\gamma^2&\gamma&0\\\gamma&1&0\\0&0&1\end{pmatrix},
 \]
where $\gamma>0$ is the amount of shear.  With $\lambda>1$ denoting the largest principal stretch it can be shown that
\[\gamma=\lambda-\lambda^{-1}. \]
We take $p=\beta$ so that $T_{33}=0$ and, because $\beta_{-1}=0$, it follows also that $T_{22}=0$.  Then the only non-zero  stresses in this simple shear are 
\be\label{eq8} T_{11}^{\rm \,ss}(\lambda)=\gamma^2\beta=(\lambda-\lambda^{-1})^2\beta,
\quad T_{12}^{\rm \,ss}(\lambda)=\gamma\beta,
\quad I_1=3+\gamma^2=\lambda^2+1+\lambda^{-2}, \en
so that $I_1$ is the same as for pure shear, as expected.  If $\lambda>1$ both these stresses are positive but if we allow $\lambda<1$ then $T_{12}^{\rm \,ss}(\lambda)$ becomes negative as $\gamma<0$.

The \emph{generalized shear modulus} is defined from Eq.\@ (\ref{eq8}) by
\be\label{eq9} \mu(\gamma^2)= T_{12}^{\rm \,ss}/\gamma = \beta-\beta_{-1}=\beta, \en
since $\beta_{-1}=0$ here,  as  there is no dependence on $I_2$.

\section{Some models for limited-stretch rubber elasticity} % section 3
\label{sec:various}

Beatty \cite{beatty2008} describes two approaches for modelling limiting polymer chain extensibility.  The first approach limits the greatest of the three principal stretches by imposing a maximum stretch $\lambda_{\rm m}$ which occurs when the polymer chains are fully extended to their maximum length. The second approach limits the value of the first principal invariant $\im$ which similarly occurs when the polymer chains are fully extended. From the experimental observations of Dickie and Smith \cite{dickie} and the theoretical results discussed by Beatty  \cite{beatty2008},    we may conclude that limiting polymer chain extensibility is governed by $\im$ alone.

We have introduced the dimensionless material constant $\im$, which is the largest value the principal invariant $I_1$ can take, and occurs when the polymer chains are  fully extended.     Therefore, $I_1$ is restricted by
\be\label{eq11}  3\leq I_1<\im. \en
Beatty \cite[Eq.\@ (7.3)]{beatty2003} showed that 
\be\label{eq10} \im=3N,   \en 
where $N$ is  the number of links forming a single polymer chain.

Apart from the neo-Hookean, all the strain energy functions we consider depend on only two material constants, a shear modulus $\mu$, and the number of links $N$ in a polymer chain, which usually appears  through $\im$, the maximum value of the first principal invariant $I_1$, see Eq.\@ (\ref{eq10}) .

For future convenience we introduce the new variable
\be\label{eq12} x=\sqrtsign{\frac{I_1}{I_m}},\quad\mbox{similarly restricted by\quad}
x_0\leq x< 1,\quad\mbox{where}\quad x_0=\sqrtsign{\frac{3}{I_m}}, \en
$x_0$ being the value of $x$ in the undeformed state, where $I_1=3$.
Let the stress response be denoted by $\beta$ when a function of $I_1$, and by $\betah$ when a function of $x$:
\[ \beta(I_1)=\betah(x). \]
From Eq.\@ (\ref{eq9}), the \emph{ground state shear modulus}, $\mu_0$, is equal to the response function $\beta$ evaluated in the undeformed state $I_1=3$ or, equivalently, $x=x_0$: 
\be\label{zz13}   \mu_0=\mu(0)=\beta(3)=\betah(x_0)=\betah(\sqrt{3/{I_m}}).   \en

If the stress response $\beta$ is known the strain energy function $W$ may be obtained by integrating Eq.\@ (\ref{eq2})$_1$
to give
\be\label{eq13}W  =     \frac12\int\beta (I_1)\,d I_1 = \im\!\int\hat{\beta}(x)\,xdx  \en
where $I_1=\im x^2$ has been used in the second integral.

\subsection{Neo-Hookean model}
\label{sec:neoH}

The neo-Hookean strain energy is given by 
\be \label{eq14}
\wnh= \tfrac12 \mu(I_1-3) = \tfrac12  \mu\im\left(x^2-\tim\right),
\en
leading to the constant response function 
\be\label{eq15}  \beta_{\rm nH} = \betah_{\rm nH}=  \mu,  \en
where $\mu_0=\mu$ is the ground state shear modulus, see Eq.\@ (\ref{zz13}).  It is the simplest possible strain energy for finite deformations in incompressible isotropic elasticity and has some degree of agreement with experiment for small to moderate strains.  The neo-Hookean model can be derived by applying Gaussian statistics to the long molecular polymer chains that make up rubber. This strain energy and stress response do not become infinite for any finite value of $I_1$ and so this is not a restricted elastic material.

\subsection{Gent's model}
\label{sec:Gent}
In Gent's model \cite[Eq.\@ (3)]{gent} the strain energy is
\be \label{eq16} \wgent =  -\frac12\mu(\im-3)\log\left( 1-\frac{I_1-3}{\im-3}\right)    \en
leading to the  response function 
\be \label{eq17}  \beta_{\rm Gent}  = \frac{\mu}{ 1-\dfrac{I_1-3}{\im-3}}
\equiv \mu \frac{\im-3}{\im - I_1} ,\quad 
 \betah_{\rm Gent}  = \left(1-\frac{3}{\im}\right)\frac{\mu}{1-x^2}.  \en
From Eqs.\@ (\ref{zz13}) and (\ref{eq17})$_1$, we see that $\mu_0=\mu$ is the ground state shear modulus.
Both strain energy and stress  become infinite as $I_1\to \im$ so that the Gent material is a restricted elastic material.  As $I_m\to\infty$, with $I_1$ remaining finite, the Gent strain energy and stress response reduce to those of the neo-Hookean material. 

\subsection{Beatty's model}
\label{sec:Beatty}

Beatty \cite[Eq.\@ (6.2)]{beatty2008} has proposed the following  model for the response function 
\be \label{eq18} \beta_{\rm \,Beatty}  = \frac{\mu}{1-\displaystyle\frac{I_1}{\im}\left(\frac{I_1-3}{\im-3}  \right)}
\equiv\frac{\mu\im (\im-3)}{(\im - I_1) (\im+I_1-3)}  ,  \en
where $\mu_0=\mu$ is the ground state shear modulus, 
leading to the strain energy
\be \label{eq19}  \wbeatty =  -\frac{\mu\im(\im-3)}{2(2\im-3)} \log\left(\dfrac{1-\dfrac{I_1-3}{\im-3}}{1+\dfrac{I_1-3}{\im}}  \right).  \en
Both strain energy and stress response  become infinite as $I_1\to \im$, so that this is a restricted elastic material.    As $\im\to\infty$, the Beatty model reduces to  the neo-Hookean model.

\subsection{Van der Waals' model}
\label{sec:Waals}

In the van der Waals model, developed by Kilian \cite{kilian}, see also  \cite[Eq.\@ (5.2)]{beatty2008} and  \cite[Eqs.\@ (24) and (25)]{horgan2002}, 
 the strain energy is given by
\be \label{eq20}
W_{\rm Waals} = -\mu(\im-3)\left[\log\left(1-\sqrt{\frac{I_1-3}{\im-3}}   \right) + \sqrt{\frac{I_1-3}{\im-3}}   \right]
\en
leading to the response function
\be \label{eq21} \beta_{\rm Waals} = \mu \left(1-\sqrt{\frac{I_1-3}{\im-3}}   \right)^{-1}
\equiv\mu\frac{\im-3+\sqrt{(I_1-3)(\im-3)}}{\im-I_1} , 
\en  
in which $\mu_0=\mu$ is the ground state shear modulus.  Once again, both strain energy and stress response  become infinite as $I_1\to \im$, so that this is a restricted elastic material.    As $\im\to\infty$, this  model also reduces to  the neo-Hookean model.  An unusual feature of this model is that the stress derivative is singular in the ground state, i.e.  as $I_1\to3$.

\subsection{Warner's model}
\label{sec:warner}

In his model for dilute suspensions of finitely extendible dumbells, Warner \cite[Eq.\@ (4)]{warner} effectively proposed the response function for nonlinear elasticity
\be\label{eq22} \betah_{\rm Warner} = \frac{\mu}{1-x^2}, \en
so that here the ground state shear modulus is given by $\mu_0=  \left(1-{3}/{I_m}\right)^{-1}\!\mu$ from Eqs.\@ (\ref{zz13}), (\ref{eq22}) and (\ref{eq12}).
The stress response (\ref{eq22}) leads to the associated strain energy
\be \label{eq23}
W_{\rm Warner} =  -\frac{1}{2} \mu\im \log\left(1-\frac{I_1-3}{I_m-3}\right).
   \en
This strain energy can be written as a multiple of Gent's strain energy (\ref{eq16}):
 \be  \label{eq24}  W_{\rm Warner} =  \left(1-{3}/{I_m}\right)^{-1} W_{\rm Gent},
   \en
an equivalence noted by Gent \cite[following Eq.\@ (6)]{gent2004}.

\subsection{FJC and WLC models}  

All the models considered in this paper are freely jointed chain (FJC) models, i.e. the polymer chain consists of rigid links that are smoothly pivoted and may have arbitrary orientation.  An alternative model is the worm-like chain (WLC) model in which the polymer chain is treated as a flexible beam that bends with temperature. Dobrynin and Carrillo
\cite{dobrynin} and Ogden et al.\@ \cite{ogden2} give excellent descriptions of the WLC model and note that it applies to biological networks and gels as well as to polymeric networks.  It is made clear in \cite{dobrynin} and \cite{ogden2} that in all FJC models the stress has a singularity like $(1-x)^{-1}$ and in all WLC models the stronger singularity $(1-x)^{-2}$.  On replacing the scaled chain length variable $x$, defined at  (\ref{eq12})$_1$,  by the first invariant $I_1$ we see that these stress singularities now behave equivalently like  $(\im-I_1)^{-1}$ and $(\im-I_1)^{-2}$, respectively.    From Eqs.\@ (\ref{eq17})$_1$, (\ref{eq18})$_2$ and (\ref{eq21})$_2$ above, the presence of the singularities $(\im-I_1)^{-1}$ confirms that Gent's, Beatty's and Van der Waals' models are FJC.  Similar analysis shows that all the models that follow have singularity $(1-x)^{-1}$ and so are FJC.

\section{The Langevin and inverse Langevin functions} % section 4
\begin{figure}[H]  % Fig. 1
\centering
\begin{tikzpicture}
\node (0,0) {\includegraphics[scale=0.75]{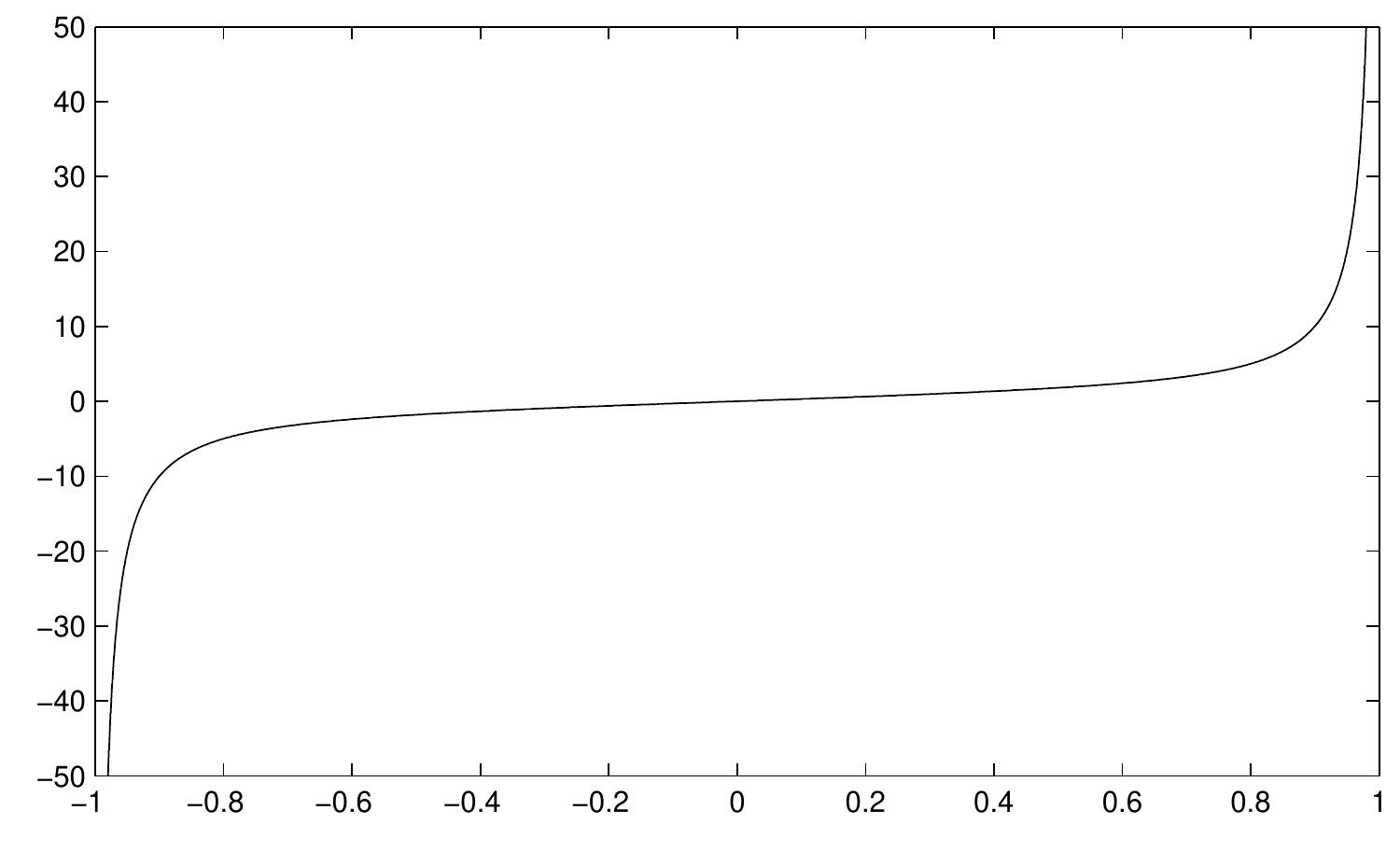}};
\draw  (-5.8,0.1) node [rotate=90] {\fontsize{10}{10} $\mathscr{L}^{-1}(x)$};
%\draw  (-4.61,3.47) node {\fontsize{10}{10} $\acute{\phantom{e}}$};
\draw  (0.3,-3.5) node {\fontsize{10}{10} $x$};
\end{tikzpicture} 
\vspace{-5pt}
\caption{Inverse Langevin function $\mathscr{L}^{-1}(x)$.  The singularities at $x=\pm1$ are apparent. They are simple poles, each with residue $-1$.}
\label{fig:1wz}
\end{figure}

\label{sec:function}
We have seen that a simple Gaussian statistics approach to rubber elasticity results in the neo-Hookean strain energy (\ref{eq14}).  A more sophisticated, non-Gaussian, statistical approach may be used to model nonlinear rubber elasticity when the limited maximum stretch of the polymer chains is taken into account.  This approach involves the inverse Langevin function.   The Langevin and inverse Langevin functions are  defined, respectively, by 
\be\label{eq25}
x=\mathscr{L}(y)=\coth y - 1/y \quad\mbox{and}\quad y=\mathscr{L}^{-1}(x)  \en
and the latter is illustrated in Figure \ref{fig:1wz}.
The Langevin function  has Taylor series %about $y=0$
\begin{equation}
\mathscr{L}(y)=\frac{1}{3}y-\frac{1}{45}y^3+\frac{2}{945}y^5-\frac{1}{4725}y^7+\frac{2}{93555}y^9-\frac{1382}{638512875}y^{11}+\cdots\,,
\label{eq26}
\end{equation}
and the inverse Langevin function  has Taylor series
\begin{equation}
\mathscr{L}^{-1}(x)=3x+\frac{9}{5}x^3+\frac{297}{175}x^5+\frac{1539}{875}x^7+\frac{126117}{67375}x^9+\frac{43733439}{21896875}x^{11}
%+\frac{231321177}{109484375}x^{13}
+\cdots\,.
\label{eq27}
\end{equation}
Itskov et al.\@ \cite{itskov2011} describe an efficient method for calculating the Taylor series for an inverse function and use it  to calculate the inverse Langevin function to 500 terms, the first 59 being presented in their paper.

\begin{figure}[H] % Fig. 2.
\centerline{
\begin{tikzpicture}
\node (0,0) {\includegraphics[scale=0.90]{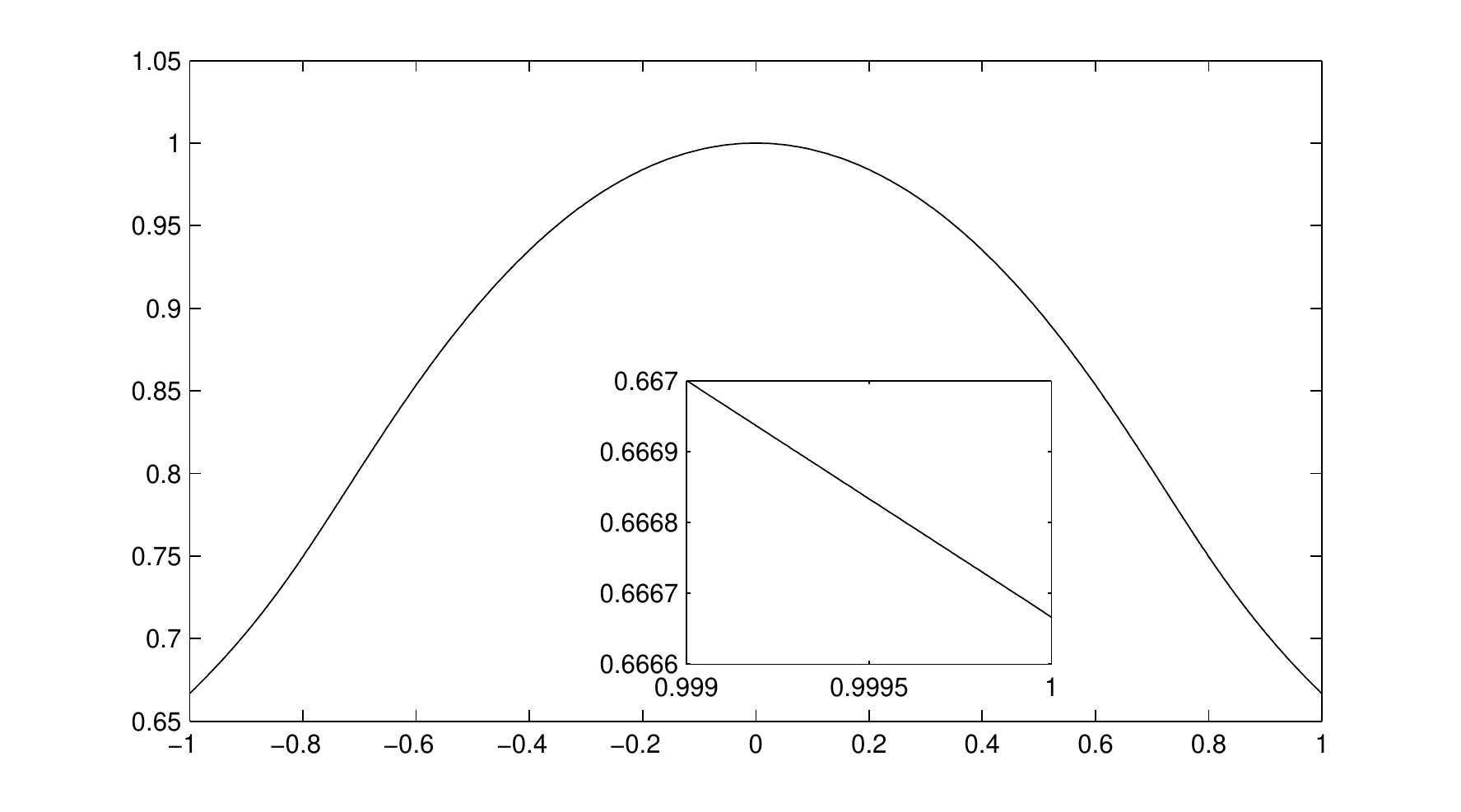}};
\draw  (-7.2,0.1) node [rotate=90] {\fontsize{10}{10} $f(x)$};
\draw  (0.3,-4.2) node {\fontsize{10}{10} $x$};
\end{tikzpicture}}  
\vspace{-5pt}
\caption{The reduced inverse Langevin function $f(x)$.   The simple poles at $x = \pm 1$ have been removed and replaced by the finite values $f(\pm1) =2/3$.}
\label{fig:2w} %Fig 2
\end{figure}

We can remove the singularities of $\mathscr{L}^{-1}(x)$ at $x=\pm1$ by considering instead the reduced inverse Langevin function $f(x)$ defined by
\begin{align}
f(x) & =\frac{(1-x^2)}{3x}\mathscr{L}^{-1}(x) \nonumber\\
 &= 
1-\frac{2}{5}x^2-\frac{6}{175}x^4+\frac{18}{875}x^6+\frac{2538}{67375}x^8+\frac{915138}{21896875}x^{10}+\cdots,
\label{eq28}
\end{align}
and  illustrated in Figure \ref{fig:2w}.

We cannot deduce the limit $\lim_{x\to\pm1} f(x)$ directly from the series expansion because this is not convergent as $x\to\pm1$.  Instead, using the fact that $x= \mathscr{L}(y) = \coth y - 1/y $, we write
\begin{align}  \lim_{x\to\pm1}\frac{1-x^2}{3x} \mathscr{L}^{-1}(x)  &= \lim_{y\to\pm\infty}\frac{1-(\coth y-1/y)^2}{3(\coth y-1/y)}\,y  \nonumber \\
  &= \lim_{y\to\pm\infty}\frac{-y/\sinh^2y+2\coth y -1/y}{3(\coth y-1/y)}   
  = \frac23,   \label{eq29} 
\end{align}
because only the terms $\coth y \to \pm1$ make a non-zero contribution to the limit.

Viewed as a function of the complex variable $x$,  the inverse Langevin function $ \mathscr{L}^{-1}(x)$ has a simple pole at $x= 1$ with residue
\be\label{eq30} \lim_{x\to 1}\, (x-1) \mathscr{L}^{-1}(x) = -1  \en
which can be deduced from the limit (\ref{eq29}).  Similarly, there is a simple pole at $x=-1$, also with residue $-1$.
These are the only real singularities of the inverse Langevin function.  The reduced inverse Langevin function $f(x)$ has no real singularities.

Itskov et al.\@ \cite{itskov2011}   calculate the radius of convergence of the power series (\ref{eq27}) to be approximately 0.904.  Since the only real singularities of  $\mathscr{L}^{-1}(x)$ are the simple poles at $x=\pm1$, this function must have further, complex, singularities within the unit circle, at a distance 0.904 from the origin.  The series (\ref{eq28}) for $f(x)$ also has  radius of convergence  0.904. 

%\subsubsection*{Arruda-Boyce eight-chain model}
\section{Limited-stretch models using the inverse Langevin function} % section 5
\label{sec:A-B}

Rubber is regarded  as being composed of cross-linked polymer chains, each chain consisting of $N$ links,   each link  having length $l$.   The distance between the two ends of the chain before deformation is denoted by $r_0$ and the distance between the two ends of the chain when fully extended is denoted by $r_{\rm L}^{\phantom{L}}$, known as the chain  locking length.   The distance between the two ends of the chain after deformation is termed the \emph{chain vector length} and is denoted by $r_{\rm chain}$.  Therefore,
\be\label{eq31} r_0\leq r_{\rm chain} \leq  r_{\rm L}^{\phantom{L}}. 
\en
The two parameters $N$ and $l$  are related to the chain vector lengths $r_{\rm L}^{\phantom{L}}$ and $r_0$ by 
\begin{equation}
 r_{\rm L}^{\phantom{L}} = Nl\quad\mbox{and}\quad r_0 = \sqrt{N}l,
\label{eq32}
\end{equation}
the former being a geometrical relationship and the latter being derived by statistical considerations.

The chain stretch is defined by $\lambda_{\rm chain}= r_{\rm chain} / r_0$  and the fully extended chain locking stretch is defined by $\lambda_{\rm L}^{\phantom{L}}= r_{\rm L}^{\phantom{L}}/r_0$.  From Eq.\@  (\ref{eq32}), we see that $\lambda_{\rm L}^{\phantom{L}}= \sqrt{N}$.   The relative chain stretch, denoted by $\lambda_\rr$, is defined to be the ratio of the current chain vector length $r_{\rm chain}$ to its fully extended length  $ r_{\rm L}^{\phantom{L}}$.  Therefore, using Eq.\@ (\ref{eq32}), we have
\be \label{eq33}
\lambda_\rr=\frac{r_{\rm chain}}{r_{\rm L}^{\phantom{L}}}= \frac{\lambda_{\rm chain}}{\lambda_{\rm L}^{\phantom{L}}} = 
\frac{\lambda_{\rm chain}}{\sqrt{N}}. 
\en
From Eqs.\@  (\ref{eq31})--(\ref{eq33}) we can deduce that $\lambda_\rr$ is restricted by
\be \label{eq34}
N^{-1/2}\leq\lambda_\rr\leq 1.
\en

\subsection{Single-chain  model}
Kuhn and Gr\"un \cite{kuhn} derived an expression for the  strain energy function of a single polymer chain.  We quote here Beatty's equivalent expression for the strain energy per unit volume, see Beatty \cite[Eqs.\@ (2.5), (2.7), (3.4) and (2.1)]{beatty2003}:
\begin{equation} 
  W(\lambda_\rr) = \mu N \left( \lambda_\rr\mathscr{L}^{-1}( \lambda_\rr)+\log \left( \frac{\mathscr{L}^{-1}( \lambda_\rr)}{\sinh \mathscr{L}^{-1}( \lambda_\rr)} \right)\right) - h_0, \label{eq35}
\end{equation}
where $\lambda_\rr$ is defined by Eq.\@ (\ref{eq33}) and  $h_0$ is a constant chosen here, and throughout, so that $W=0$ in the reference configuration.     See also Wu and van der Giessen \cite[Eq.\@ (1)]{wu}.

\subsection{Three-chain  model}
The original three-chain model of James and Guth \cite{james} is based on three independent polymer chains, each with the same initial chain vector length, and each parallel to an axis of an orthogonal  Cartesian coordinate system. The James and Guth \cite{james} strain energy for this three-chain system  may be written as
\begin{equation} 
  \wJG = \frac{\mu N_3}{3} \sum^3_{j=1}\left( \alpha_j\mathscr{L}^{-1}(\alpha_j)+\log \left( \frac{\mathscr{L}^{-1}(\alpha_j)}{\sinh \mathscr{L}^{-1}(\alpha_j)} \right)\right) - h_0 \label{eq36}
\end{equation}
where 
\be\label{eq37} \alpha_j={\frac{\lambda_j}{\sqrt{N_3}}},
\en
see Beatty \cite[Eq. (3.2)]{beatty2003}.
The $\lambda_j$, for $j=1, 2, 3$, denote the principal stretches along the coordinate axes. 
The parameter $N_3$ is the number of links in each chain of the three-chain model and, as such, is formally the same as the parameter $N$ occurring in other models.  However, to achieve  good agreement between the three-chain model and the eight-chain model, discussed below, we have to allow $N_3$ and $N$ to be different.  This is discussed in more detail in Section \ref{sec:ABJG}.

The strain energy (\ref{eq36}) has been multiplied by a factor of $1/3$ to ensure the three-chain network model has same total entropy as the single-chain network model, see Treloar \cite[page 114]{treloar1975} for further discussion.

If we approximate $\mathscr{L}^{-1}(\alpha_j)\approx 3\alpha_j+9\alpha_j^3/5$, from the series (\ref{eq27}),  in Eq.\@  (\ref{eq36})  we find that in the limit $N_3\to\infty$ the strain energy (\ref{eq36}) becomes the neo-Hookean strain energy, as expected.  

For finite $N_3$, the strain energy (\ref{eq36}) becomes infinite as $\alpha_j\to 1$, i.e., as $\lambda_j\to \sqrt{N_3}$, so that the three-chain model is a limited-stretch model of rubber elasticity.   Each principal stretch $\lambda_j$ is separately limited by the value $\sqrt{N_3}$.

Because Eq.\@ (\ref{eq36}) is a symmetric function of the principal stretches, and therefore represents an isotropic elastic material, we may use  the strain energy (\ref{eq36}) in Eq.\@ (\ref{eq4}) to give the principal stresses for the three-chain model:
\be\label{eq38}
T_j= -p + \frac{\mu N_3}{3} \alpha_j\mathscr{L}^{-1}(\alpha_j),\quad\mbox{for}\quad j=1,2,3,
\en
see  Beatty \cite[Eq. (3.3)]{beatty2003} and also
 Wu and van der Giessen \cite[Eq.\@ (30)]{wu}, James and Guth \cite[Eq.\@ (6.9)]{james} and Wang and Guth \cite[Eq.\@ (4.13a)]{wang}.  In the last two references we must assign the authors' parameter $\kappa$ the value $\kappa=1/\sqrt{N_3}$ in order to obtain the agreement of their stresses with Eq.\@ (\ref{eq38}).

\subsection{Eight-chain  model}
 Consider a cube of side $2a$ centred on the origin with edges parallel to the coordinate axes.  In the Arruda-Boyce \cite{arruda} eight-chain  model of rubber elasticity,  each of the eight vertices $(\pm a, \pm a, \pm a)$ of the cube is joined to the origin by a polymer chain.   We suppose that each of the eight identical chains has  initial  chain vector length $r_0$, given by Eq.\@ (\ref{eq32})$_2$, so that by geometry the  length $a$ satisfies
\be \label{eq39}  \sqrt{3}a= r_0=\sqrt{N}l. \en
If the deformation of the elastic material is triaxial, with principal stretches $\lambda_j$ for $j=1,2,3$ in directions parallel to the edges of the cube, then the cube deforms into a cuboid  with sides 
$2\lambda_1a$, $2\lambda_2a$ and $2\lambda_3a$.   All eight chains now have the same chain vector  length $r_{\rm chain}$ given by
\begin{align} r_{\rm chain} &= \sqrt{(\lambda_1a)^2+(\lambda_2a)^2+(\lambda_3a)^2}  \nonumber \\
 &= \sqrt{\lambda_1^2+\lambda_2^2+\lambda_3^2}\,a  \nonumber \\
 &=\sqrt{ \frac{I_1N}{3}}l, \label{eq40}
 \end{align}
 using Eqs.\@ (\ref{zz5}) and (\ref{eq39}).
The relative chain stretch $ \lambda_{\rm r}$ and the chain stretch $\lambda_{\rm chain}$  are given by
\begin{equation} \lambda_{\rm r}= \frac{r_{\rm chain}}{r_{\rm L}^{\phantom{L}}} =\sqrtsign{\frac{I_1}{I_m}}=x
\quad\mbox{and}\quad \lambda_{\rm chain} =     \sqrtsign{\frac{I_1}{3}},
\label{eq41}
\end{equation}
respectively, for the Arruda-Boyce model, where Eqs.\@ (\ref{eq33}), (\ref{eq40}),  (\ref{eq32})$_1$,  (\ref{eq10}) and  (\ref{eq12})$_1$ have been used.

The Arruda-Boyce eight-chain stress response function is given by 
\be\label{eq42}
\betah_{\rm 8ch} = \mu \mathscr{L}^{-1}(x)/3x.   
\en
Using Eq.\@ (\ref{eq13}), we can integrate $\betah_{\rm 8ch}$ above in order to find the strain energy
\begin{align}
\wab  
 & = \frac{ \mu\im}{3}\int\mathscr{L}^{-1}(x)\,dx \nonumber  \\
 & = \frac{ \mu\im}{3} \left( x\mathscr{L}^{-1}(x)+\log \left( \frac{\mathscr{L}^{-1}(x)}{\sinh \mathscr{L}^{-1}(x)} \right)\right) - h_0. \label{eq43}
\end{align}
The integration is carried out using the general formula  for the integral of an inverse function $y=f^{-1}(x)$,
easily proved by integration by parts:
\[ \int f^{-1}(x)\,dx = xf^{-1}(x) - \int f(y)\,dy,  \]
see, for example, Parker \cite{parker}.

On approximating $\mathscr{L}^{-1}(x)\approx 3x+9x^3/5$ from the series (\ref{eq27})  in Eq.\@  (\ref{eq43}),  we find that as $\im\to\infty$ the strain energy (\ref{eq43}) becomes the neo-Hookean strain energy, as expected.  
 
The stress response  (\ref{eq42}) and strain energy (\ref{eq43}) become infinite as $x\to 1$, i.e., as $I_1\to \im$, for finite $\im$, so that the eight-chain model is a limited-stretch model of rubber elasticity.

 From Eqs.\@ (\ref{zz13}) and (\ref{eq42}), the ground state shear modulus is $\mu_0=\mu\,\mathscr{L}^{-1}(x_0)/3x_0$ where $x_0$ is the ground state value of $x$, defined at Eq.\@ (\ref{eq12}).  For the value $\im=60$ employed in our numerical work later we find that $\mu_0$ is close to $\mu$; in fact,       $\mu_0\approx 1.03\mu$.
%  Actually,  $\mu_0\approx 1.031491695\mu$.

There is a close connection between the single-chain and eight-chain models:
replacing $\lambda_\rr$  in the single-chain strain energy (\ref{eq35}) by $x$, from  Eq.\@ (\ref{eq41}),  results immediately in the eight-chain strain energy (\ref{eq43}), having used Eq.\@  (\ref{eq10}).

 Beatty \cite[Eq.\@ (6.4)]{beatty2003}, and earlier Dickie and Smith  \cite[Eq. (30)]{dickie}, showed that the average chain stretch of a randomly oriented molecular chain is given by  Eq.\@ (\ref{eq41})$_2$, the same as for the Arruda-Boyce model.
 Beatty \cite[Section 6]{beatty2003} demonstrated the remarkable result that the Arruda-Boyce stress response Eq.\@ (\ref{eq42}) holds in general for an average stretch, full-network model of arbitrarily oriented molecular chains.  Therefore, the eight-chain cell structure is unnecessary.

%\clearpage
\subsection{Four-chain  model}
The Wang and Guth \cite{wang}  four-chain model of rubber elasticity is based on a regular tetrahedron in which each vertex is joined by a polymer chain to the centre of the tetrahedron.  However, it is a fact that a regular tetrahedron can be embedded in a cube with its four vertices coinciding with four of the vertices of the cube.  For example, the four vertices $(a,a,a)$, $(a,-a,-a)$, 
$(-a,a,-a)$, $(-a,-a,a)$ of the Arruda-Boyce cube of side $2a$ form a regular tetrahedron with each edge of length $2\sqrt{2}a$.  The centre of this tetrahedron coincides with the centre of the cube, at the origin.  Therefore, the four chains that link the four vertices of this tetrahedron to its centre coincide with four of the eight chains of the eight-chain model.  It follows that the four-chain model is entirely equivalent to the eight-chain model, as stated by Beatty \cite{beatty2003} and {Z\'{u}\~{n}iga and Beatty  \cite{zuniga2002}.

Because of the equivalence of the four-chain and eight-chain models, the four-chain stress response is identical to the eight-chain stress response, given by Eq.\@ (\ref{eq42}), and the four-chain strain energy is identical to the eight-chain strain energy, given by Eq.\@ (\ref{eq43}).      Wang and Guth \cite[Eq.\@ (4.13b)]{wang} obtain the stress  Eq.\@ (\ref{eq42}) in their four-chain model provided we    assign their parameter $\kappa$ the value $\kappa=1/\sqrt{3N}$.   This observation has been made also by {Z\'{u}\~{n}iga and Beatty  \cite[Eq.\@ (5.3)]{zuniga2002}.

\subsection{Numerical comparison of the  three-chain model with the  eight-chain model} % subsection  5.1
\label{sec:ABJG}
We have already mentioned the fact that in order to obtain good agreement between the three-chain and eight-chain models we have to permit the number of links per chain $N_3$ in the three-chain model to be different from $N$, the number of links per chain in the eight-chain model.  In order to obtain the closest possible fit between the three-chain strain energy (\ref{eq36}) and the eight-chain strain energy (\ref{eq43}) we must ensure that the singularities of the dominant inverse Langevin function of Eq.\@  (\ref{eq36})  and the inverse Langevin function of Eq.\@  (\ref{eq43}) both occur at the same stretch, which is the maximum possible stretch.   

We now derive connections between $N_3$ and $N$ for various types of deformation,  see Z\'{u}\~{n}iga and Beatty \cite[Appendix A.2]{zuniga2002} for uniaxial tension and compression. 

\subsubsection{Uniaxial tension and compression}
\begin{figure}[h] % Fig. 3
    \centering
\subfigure [Uniaxial tension  and compression $(\lambda\gtrless 1)$]{
    \begin{tikzpicture}
\node[anchor=north west,inner sep=-2] at (0,0) {\includegraphics[scale=0.85]{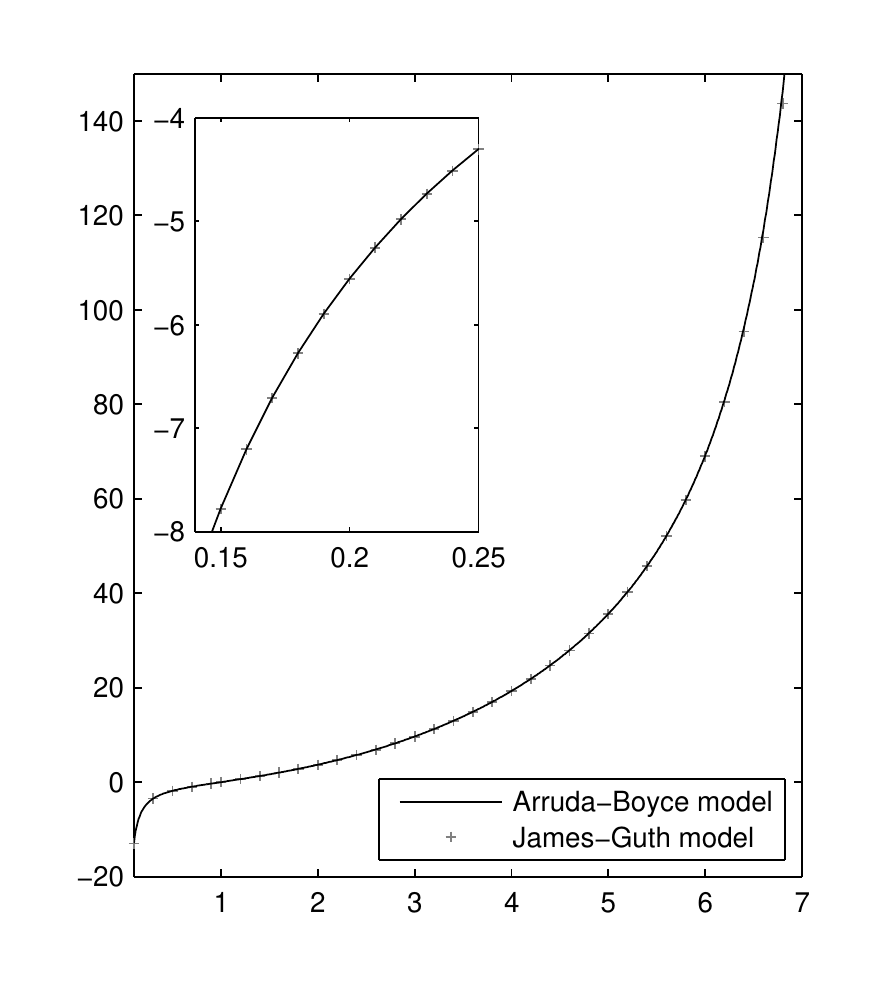} };    
\draw  (0.2,-4.1) node [rotate=90] {\fontsize{10}{10} $T^{\rm{uni}}_{11}(\lambda)/\mu$};
\draw  (3.9,-8.1) node {\fontsize{10}{10} $\lambda$};
\end{tikzpicture}} 
%    \qquad
    \subfigure [Biaxial tension and compression $(\lambda\gtrless 1)$]{
    \begin{tikzpicture}
\node[anchor=north east,inner sep=-2] at (0,0){\includegraphics[scale=0.85]{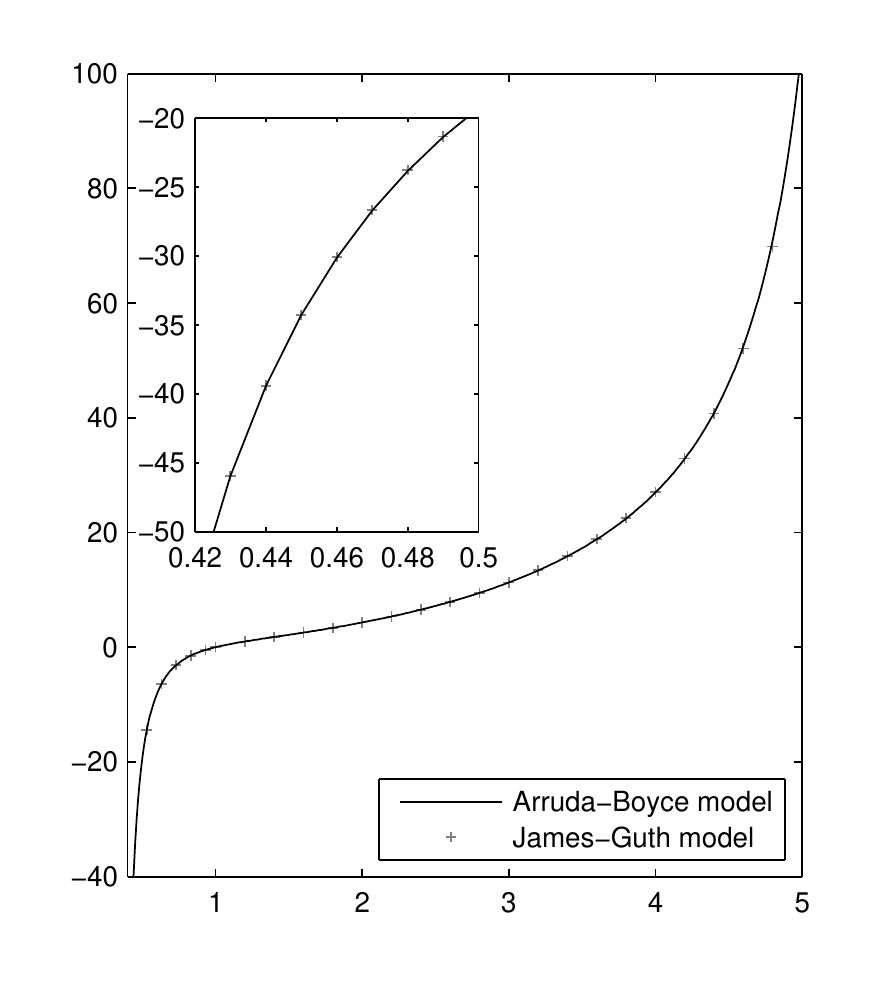} };
\draw  (-7.3,-4.1) node [rotate=90] {\fontsize{10}{10} $T^{\rm{bi}}_{11}(\lambda)/\mu$};
\draw  (-3.6,-8.1) node {\fontsize{10}{10} $\lambda$};
\end{tikzpicture}}  
    \caption{Comparison between the Arruda-Boyce eight-chain model and the James-Guth three-chain model. For uniaxial tension and biaxial tension in the Arruda-Boyce model we have taken $N=20$. In James-Guth model: for uniaxial tension $N_3=59.7$ from Eq.\@  (\ref{eq45}), for uniaxial compression $N_3=30.0$ from Eq.\@  (\ref{eq46}), for biaxial extension $N_3=30.0$ from Eq.\@  (\ref{eq46}), for biaxial compression $N_3=59.7$ from Eq.\@  (\ref{eq45}).}%
    \label{fig:JG1}%    
\end{figure}
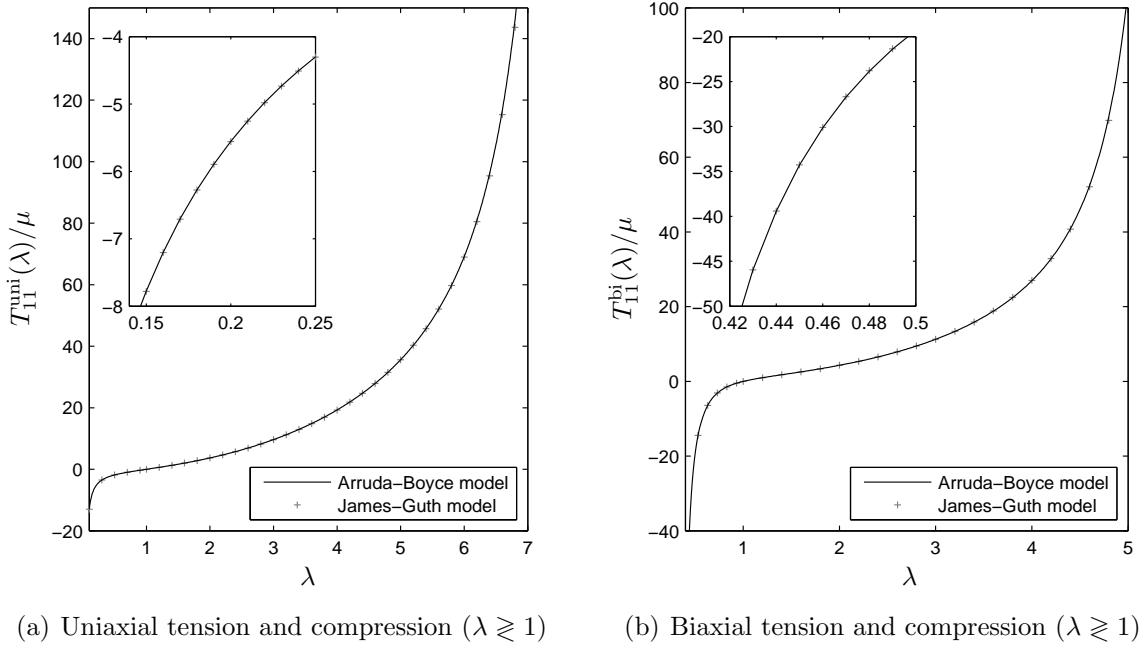

In uniaxial tension the largest stretch is $\lambda_1=\lambda>1$ with the smaller stretches given by $\lambda_2=\lambda_3=1/\sqrt{\lambda}$.
Taking $\alpha_j=\lambda$, the greatest stretch, in   Eq.\@  (\ref{eq37}) and  $x=\sqrt{I_1/\im}$, from Eq.\@  (\ref{eq12})$_1$, the theoretical best fit between the two models occurs when  
\be\label{eq44}
 \frac{\lambda}{\sqrt{N_3}} = \sqrt{\frac{I_1}{\im}} =  \sqrt{\frac{\lambda^2+2\lambda^{-1}}{3N}} = 1, 
 \en 
 where Eqs.\@ (\ref{eq5})$_2$ and  (\ref{eq10}) have also been used.
We see that $\lambda=\sqrt{N_3}=\lambda_{\rm L}$, the chain locking stretch, is the maximum possible stretch.
From the last equation of Eq.\@ (\ref{eq44}), we obtain
 a relation connecting the three-chain  and eight-chain models in uniaxial tension:
\begin{equation} 3N= N_3+\frac{2}{\sqrt{N_3}}.
\label{eq45}
\end{equation} 
In later numerical work we take $\im=60$ so that $N=20$ from Eq.\@ (\ref{eq10}).
From  Eq.\@ (\ref{eq45}), we see that  $N_3\approx 59.7$ if $N=20$.

In a uniaxial compression, with uniaxial stretch $\lambda_1=\lambda<1$, the greatest  stretches are 
 $\lambda_2=\lambda_3=1/\sqrt{\lambda}$, in the lateral directions.   In this case Eq.\@  (\ref{eq44}) 
is replaced by
\[ \frac{1/\sqrt{\lambda}}{\sqrt{N_3}} = \sqrt{\frac{I_1}{\im}} =  \sqrt{\frac{\lambda^2+2\lambda^{-1}}{3N}} = 1, \]
so that  now $\lambda=1/{N_3}$.   
The largest possible stretch is $\lambda_2=\lambda_3=1/\sqrt{\lambda}=N_3^{1/2}=\lambda_{\rm L}$, the chain locking stretch, as for uniaxial tension.
Arguing as before we obtain
 a relation connecting the three-chain  and eight-chain models, this time in uniaxial compression:
 \begin{equation} 3N= 2N_3+\frac{1}{{N_3^2}},
\label{eq46}
\end{equation} 
so that now   $N_3\approx 30$ if $N=20$.  
See Figure \ref{fig:JG1}(a). 

\subsubsection{Biaxial tension and compression}

For a biaxial deformation $I_1=2\lambda^2+\lambda^{-4}$, see Eq.\@ (\ref{eq6})$_2$. 
In  biaxial tension the two largest stretches are $\lambda_1=\lambda_2=\lambda$.   In this case Eq.\@  (\ref{eq44}) 
is replaced by
\[ \frac{\lambda}{\sqrt{N_3}} = \sqrt{\frac{I_1}{\im}} =  \sqrt{\frac{2\lambda^2+\lambda^{-4}}{3N}} = 1, \]
so that  now $\lambda=\sqrt{N_3}=\lambda_{\rm L}$.   
Arguing as before  leads to the relation
\[ 3N= 2N_3+\frac{1}{N_3^2},
\label{jg3}
\] 
previously obtained for uniaxial compression, see Eq.\@ (\ref{eq46}).
As  for uniaxial compression, $N_3\approx30$ if $N=20$.   Thus, biaxial tension is similar to uniaxial compression.

Biaxial compression, however, is similar to uniaxial tension.   The greatest  stretch is now $\lambda_3=1/{\lambda^2}$ and Eq.\@  (\ref{eq44}) 
is replaced by
\[ \frac{1/\lambda^2}{\sqrt{N_3}} = \sqrt{\frac{I_1}{\im}} =  \sqrt{\frac{2\lambda^2+\lambda^{-4}}{3N}} = 1, \]
so that  now $\lambda= N_3^{-1/4}$.   The greatest stretch remains
 $\lambda_3=\sqrt{N_3}=\lambda_{\rm L}$.
Arguing as before  leads to the relation
\[ 3N= N_3+\frac{2}{\sqrt{N_3}},
\label{jg4}
\]
for biaxial compression, previously obtained for uniaxial tension, see Eq.\@ (\ref{eq45}).
As for uniaxial tension, $N_3\approx59.7$ if $N=20$.   See Figure \ref{fig:JG1}(b).

\subsubsection{Pure shear}

For  pure shear $I_1=\lambda^2+1+\lambda^{-2}$, see Eq.\@ (\ref{eq7})$_2$. 
In a pure shear extension the greatest stretch is $\lambda_1=\lambda>1$.  In this case Eq.\@ (\ref{eq44})   is replaced by
\[
 \frac{\lambda}{\sqrt{N_3}} = \sqrt{\frac{I_1}{\im}} =  \sqrt{\frac{\lambda^2+1+\lambda^{-2}}{3N}} = 1, 
 \]
Arguing as before gives  $\lambda={N_3^{1/2}}$ and leads to the relation:
\begin{equation} 3N= N_3+1+\frac{1}{N_3},
\label{eq47}
\end{equation} 
so that $N_3\approx59.0$ if $N=20$.

In a pure shear compression    $\lambda_1=\lambda<1$ is the smallest stretch and the largest is    
$\lambda_3=1/{\lambda}$.  Arguing as before gives {$\lambda=N_3^{-1/2}$ and leads once more to the relation  (\ref{eq47}).   See Figure \ref{fig:JG2}(a).

\subsubsection{Simple shear}

In simple shear we have seen that the amount of shear  $\gamma$ and the largest principal stretch $\lambda$ satisfy the relation $\gamma=\lambda-\lambda^{-1}$, which can be solved for $\lambda$: 
\[ \lambda = \frac12\gamma+\frac12\sqrt{\gamma^2+4}.   \]
In this case Eq.\@ (\ref{eq44}) is replaced by
\be \label{eq48}   
\frac{ \frac12\gamma+\frac12\sqrt{\gamma^2+4}}{\sqrt{N_3}} = \sqrt{\frac{I_1}{\im}} =  \sqrt{\frac{3+\gamma^2}{3N}} = 1,   \en
where Eq.\@ (\ref{eq8})$_3$ has been used.  Arguing as before gives
\[  \frac12\gamma+\frac12\sqrt{\gamma^2+4} = \sqrt{N_3} \implies \gamma=\sqrt{N_3}-1/\sqrt{N_3}.  \]
Using this last relation in the last of Eqs.\@  (\ref{eq48}) then leads to  Eq.\@ (\ref{eq47}) in simple shear, the same as for pure shear.
This is not surprising as simple shear is simply a rotation of  pure shear.     See Figure \ref{fig:JG2}(b).

Figures \ref{fig:JG1} and \ref{fig:JG2} demonstrate the excellent correlation between the Arruda-Boyce and the James-Guth models.   However, the Arruda-Boyce model is easier to implement and   the number of links per polymer chain $N$ is the same for all deformations, whereas for the James-Guth model  uniaxial and biaxial deformations require a different number of links $N_3$ in compression and extension.  This would not be expected experimentally.

 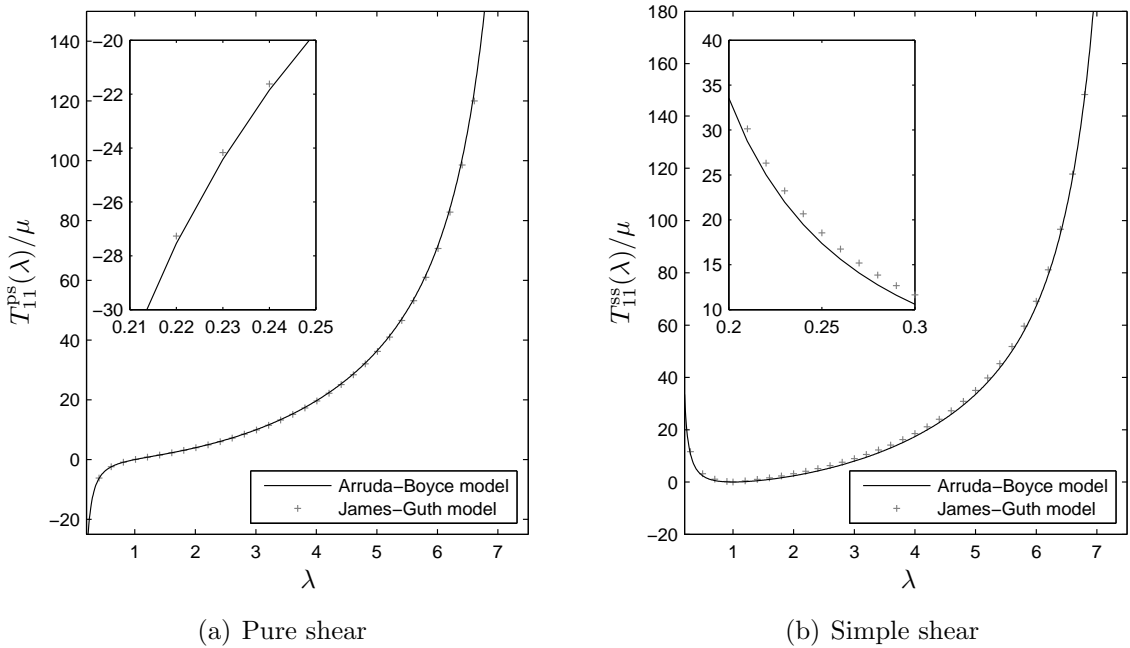
\begin{figure}[H]  % Fig. 4
    \centering
\subfigure [Pure shear]{
    \begin{tikzpicture}
\node[anchor=north west,inner sep=-2] at (0,0) {\includegraphics[scale=0.85]{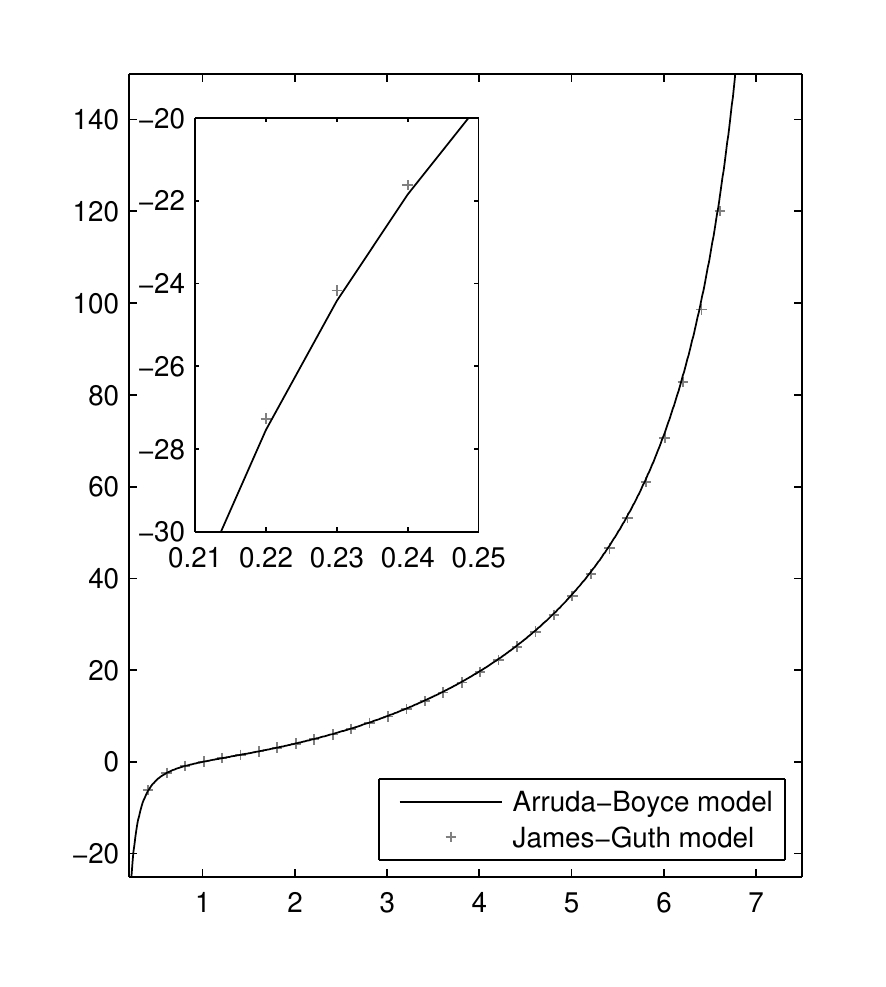} };    
\draw  (0.2,-4.1) node [rotate=90] {\fontsize{10}{10} $T^{\rm{ps}}_{11}(\lambda)/\mu$};
\draw  (3.9,-8.1) node {\fontsize{10}{10} $\lambda$};
\end{tikzpicture}}    
 %    \qquad
    \subfigure [Simple shear]{
    \begin{tikzpicture}
\node[anchor=north east,inner sep=-2] at (0,0){\includegraphics[scale=0.85]{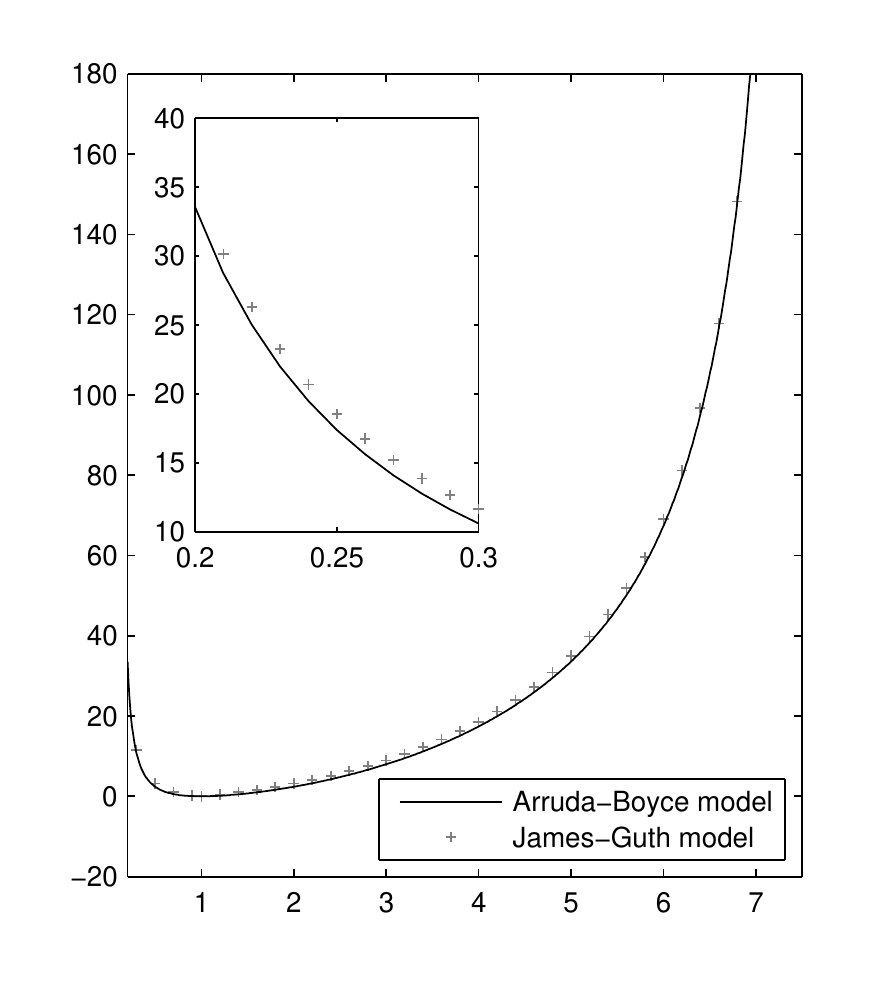} };
\draw  (-7.3,-4.1) node [rotate=90] {\fontsize{10}{10} $T^{\rm{ss}}_{11}(\lambda)/\mu$};
\draw  (-3.6,-8.1) node {\fontsize{10}{10} $\lambda$};
\end{tikzpicture}}  
      \caption{Comparison between the Arruda-Boyce eight-chain model and the James-Guth three-chain model. For both pure  and simple shear in the Arruda-Boyce model we have taken $N=20$. In the James-Guth model, for both these shear deformations we have taken $N_3=59.0$ from Eq.\@  (\ref{eq47}).     }%
    \label{fig:JG2}%    
\end{figure}

\section{Models based on approximations to the inverse Lang\-evin function} % section 6
\label{sec:approcinvlan}

\subsection{Power series methods}

Our first approximations to the inverse Langevin function are obtained simply by truncating its power series expansion (\ref{eq27}), see for example \cite{bol} and \cite{chagnon2006}.   Itskov et al.\@ \cite{itskov2011} estimate the radius of convergence of the power series to be $0.904$.  Because this radius of convergence is less than unity, the power series is unable to capture correctly the behaviour close to the singularities at $x= \pm 1$, as illustrated in Figure \ref{fig:1w}.  

All the models which follow are based on approximations to the inverse Langevin function.  Many of them employ Pad\'e approximants which are defined as follows.  The $[M/N]$ Pad\'e approximant to a function $F(x)$ in the neighbourhood of the origin is the (unique) rational function $P_M(x)/Q_N(x)$, where $P_M(x)$ is a polynomial of degree $M$ in $x$ and $Q_N(x)$ is a polynomial of degree $N$ in $x$ (normalized so that $Q_N(0)=1$), chosen so that
\[ F(x)-\frac{P_M(x)}{Q_N(x)}=O(x^{M+N+1}), \]
see, for example, Hinch \cite[pp 152--153]{hinch}.  In other words, the coefficients of $P_M(x)$ and $Q_N(x)$ must be chosen so that the Taylor series about $x=0$ of $P_M(x)/Q_N(x)$ matches exactly that of $F(x)$ up to, and including, the term in $x^{M+N}$.

\begin{figure}[H] % Fig. 5
\centering
\begin{tikzpicture}
\node (0,0) {\includegraphics[scale=0.75]{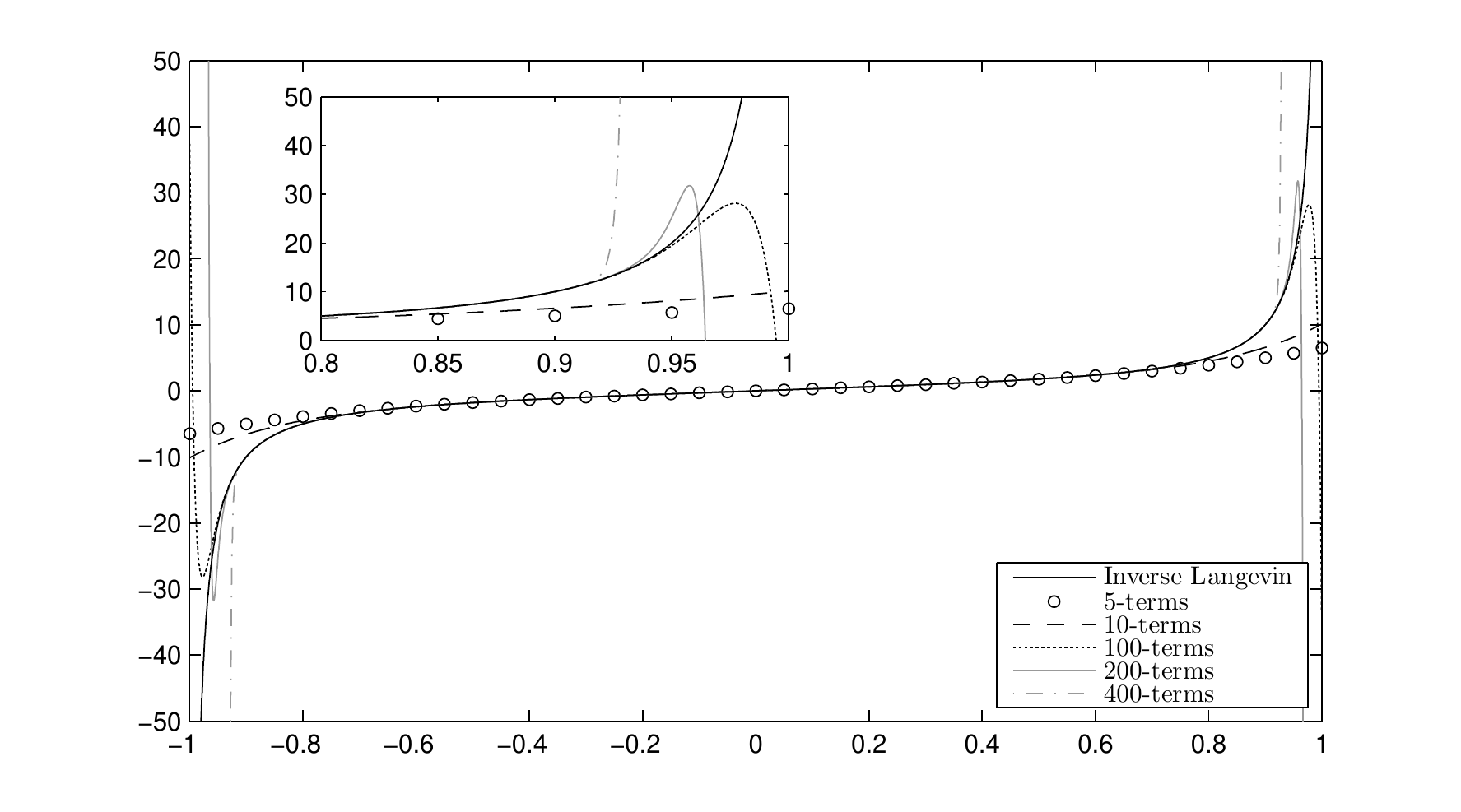}};
\draw  (-5.8,0.1) node [rotate=90] {\fontsize{10}{10} $\mathscr{L}^{-1}(x)$};
%\draw  (-4.61,3.47) node {\fontsize{10}{10} $\acute{\phantom{e}}$};
\draw  (0.3,-3.5) node {\fontsize{10}{10} $x$};
\end{tikzpicture} 
\vspace{-5pt}
\caption{Inverse Langevin function $\mathscr{L}^{-1}(x)$ with simple poles at $x=\pm1$.  Shown also are various truncated power series approximations. }
\label{fig:1w}
\end{figure}

Horgan and Saccomandi \cite[Eq.\@ (19)]{horgan2002} introduce a model consisting of the $[1/2]$ Pad\'e approximant of the inverse Langevin function,
\begin{equation}
 \mathscr{L}^{-1}(x) \approx \frac{3x}{1-\frac35x^2},
\label{x37}
\end{equation}
but we do not pursue this further as it does not have the required singularities at $x=\pm1$, as noted in \cite{horgan2002}.

\subsection{Cohen's model}
\label{sec:Cohen}
Cohen \cite{cohen} showed that the $[3/2]$ Pad\'e approximant of $\mathscr{L}^{-1}(x) $ is
\[ \mathscr{L}^{-1}(x)  \approx x\frac{3-\frac{36}{35}x^2}{1-\frac{33}{35}x^2}. \]
This does not have the required simple poles at $x=\pm1$ and so Cohen further approximated by replacing each fraction by 1:
\be \label{e13}   \mathscr{L}^{-1}(x) \approx  3x \,\frac{1-\frac13 x^2}{1-x^2}.   \en
This approximation happens to have the same real singularities as the inverse Langevin function, namely, simple poles at $x=\pm1$, each with residue $-1$.
With the approximation (\ref{e13}), the  response function  is 
\be \label{e14}  \betah_{\rm Cohen} = \mu  \frac{1-\frac13 x^2}{1-x^2},  \en
so that $\mu_0\approx 1.035\mu$  from Eq.\@ (\ref{zz13}) with $\im=60$. 
From (\ref{eq13}),  the  strain energy is given by
\begin{align}
\wcohen  
  &=  \mu\im\int \frac{1-\frac13 x^2}{1-x^2} \,xdx  \nonumber\\
  &=  \frac12\mu\im\left(\frac13 x^2-\frac23 \log(1-x^2)\right) -h_0  \nonumber\\
  &=  \frac12\mu\left(\frac13 I_1-\frac23\im\log\left(1-\iim\right) \right) -h_0.  \nonumber\\
\intertext{This gives, for suitable choice of $h_0$,}
\wcohen &=  \frac16\mu(I_1-3) - \frac13 \im  \log\left( \frac{\displaystyle1-\iim}{\displaystyle1-\tim} \right), \nonumber\\
\intertext{which vanishes in the reference state $I_1=3$.  This can be written, using the identity}
 \left(\displaystyle1-\iim\right)\left/\left(\displaystyle1-\tim \right) \right. &=  1-\frac{I_1-3}{\im-3}   , \label{e15}  \\
\intertext{as}
\wcohen &=  \frac16\mu(I_1-3) - \frac13 \im  \log\left(1-\frac{I_1-3}{\im-3}  \right). \nonumber\\
\intertext{From the definition (\ref{eq16}) of the Gent strain energy we see that this becomes}
\wcohen &=  \frac13\wnh + \frac23(1-3/\im)^{-1}\wgent, \label{e16}
\end{align}
so that Cohen's form of the strain energy is simply a linear combination of the neo-Hookean and Gent forms.  It follows that Cohen's stress response is the same linear combination of the neo-Hookean and Gent stress responses:
\be\label{e16a}  \betah_{\rm Cohen} =  \frac13\beta_{\rm nH} + \frac23(1-3/\im)^{-1}\beta_{\rm Gent}.
\en

\subsection{New model}
\label{sec:New}

We wish to restrict attention to those approximations of $\invL$ which have  simple poles at $x=\pm1$.  Therefore, we approximate instead of $\invL$, given by Eq.\@ (\ref{eq27}),  the reduced inverse Langevin function $f(x)$ defined by  Eq.\@ (\ref{eq28}).  On taking the first two terms of the series in  Eq.\@ (\ref{eq28}) we obtain the  approximation
\be \label{e17} \mathscr{L}^{-1}(x) \approx  3x \,\frac{1-\frac25 x^2}{1-x^2},   \en
which is very similar to Cohen's approximation (\ref{e13}).  This approximation has simple poles at $x=\pm1$ but with residues $-9/10$ instead of the correct $-1$.  Using the approximation (\ref{e17}) gives rise to a new model with  response function 
\be \label{e18}  \betah_{\rm New} = \mu  \frac{1-\frac25 x^2}{1-x^2},   \en
so that $\mu_0\approx 1.032\mu$  from Eq.\@ (\ref{zz13}) with $\im=60$.   The  strain energy is
\be \label{e19} \wnew =  \frac25\wnh + \frac35(1-3/\im)^{-1}\wgent, \en
also a linear combination of the neo-Hookean and Gent forms.   It follows that the stress response of new model is the same linear combination of the neo-Hookean and Gent stress responses:
\be\label{e16b}  \betah_{\rm New} =  \frac25\betah_{\rm nH} + \frac35(1-3/\im)^{-1}\betah_{\rm Gent}.
\en
We shall see that this new model is a very good approximation to the inverse Langevin model and so  Eqs.\@  (\ref{e19}) and (\ref{e16b}) are evidence of a very close empirical connection between the inverse Langevin  model and the much simpler neo-Hookean and Gent models.

The new approximation (\ref{e17}) is, in fact, based on the [2/0] Pad\'e approximant of $f(x)$.  We have also investigated models based on the [4/0] and [2/2] Pad\'e approximants of $f(x)$ but found them to be no more accurate than (\ref{e17}).
Below, we shall see that the [0/4] Pad\'e approximant of $f(x)$ leads to a model,  the modified Treloar model,   even more accurate than~(\ref{e17}).

\subsection{Treloar's model}
\label{sec:Treloar}

Treloar \cite[ Eq.\@  (9.6a)]{treloar1975} has approximated the inverse Langevin function $ \mathscr{L}^{-1}(x)$ by its [1/6] Pad\'{e} approximant to give
\be  \label{e22}   \mathscr{L}^{-1}(x) \approx  \frac{3x}{1-\frac35x^2-\frac{36}{175}x^4-\frac{108}{875}x^6}.  \en
This does not have the required singularities at $x=\pm 1$ but Treloar's further approximation to this \cite[ Eq.\@  (9.6d)]{treloar1975}, namely,
\be   \label{e23}  \mathscr{L}^{-1}(x) \approx  \frac{3x}{1-\frac35x^2-\frac15x^4-\frac15x^6} = 
 \frac{3x}{(1-x^2)(1+\frac25x^2+\frac15x^4)} \en
does have the required singularities\footnote{Simple poles at $x=\pm1$ each with residue $-15/16 = -0.9375$ instead of $-1$.}.  It gives rise to the stress response function
\be\label{e24} \betah_{\rm \,Treloar}= \frac{\mu}{(1-x^2)(1+\frac25x^2+\frac15x^4)}, \en
so that $\mu_0\approx 1.031\mu$  from Eq.\@ (\ref{zz13}) with $\im=60$. 
The strain energy is
\begin{align}
W_{\rm Treloar}
     &=  \frac{5}{32}\mu\im\left[ \log\left(\frac{1+\frac25x^2+\frac15x^4}{(1-x^2)^2}\right) +2\arctan \left(\frac{1+x^2}{2} \right)  \right]-h_0.  \label{e25a} 
\end{align}

\subsection{Modified Treloar model}
\label{sec:modTreloar}

We wish to modify Treloar's model above so that it treats the singularities at $x=\pm 1$ exactly.
If we wish to go as far as terms in $x^6$ in the denominators, then taking the $[1/6]$ Pad\'e approximant of $\invL$ given by  (\ref{eq27}) is equivalent to taking the $[0/4]$ Pad\'e approximant of $f(x)$ given by  (\ref{eq28}), except that the singularities at $x=\pm1$ are now built in.
The $[0/4]$ Pad\'{e} approximant of $f(x)$ is
\[ f(x) \approx \frac{1}{1+\frac25x^2+\frac{34}{175}x^4}, \]
leading to the approximation\footnote{Simple poles at $x=\pm1$ each with residue $- 525/558\approx - 0.9409$ instead of $-1$.}
\be  \label{e26}  \mathscr{L}^{-1}(x) \approx  \frac{3x}{(1-x^2)(1+\frac25x^2+\frac{34}{175}x^4)}\en
which differs from Treloar's (\ref{e23})$_2$ by only 1/175 in  the coefficient of $x^4$.  The corresponding  stress response  is 
\be
\label{a27} \betah_{\rm mod\underline{\;\;}\!Treloar} = 
\frac{\mu}{(1-x^2)(1+\frac25x^2+\frac{34}{175}x^4)}, 
\en
so that now $\mu_0\approx 1.032\mu$ with $\im=60$,  
and the corresponding strain energy is
%\be 
\begin{multline}\label{e27}  W_{\rm mod\underline{\;\;}\!Treloar} = 
 \frac{5}{31} \mu\im \left[  \frac{35}{36}\log\left(\frac{1+\frac25x^2+\frac{34}{175}x^4}{(1-x^2)^2}\right)
   + \frac{23\sqrt{21}}{54}\arctan\left(\frac{1+\frac{34}{35}x^2}{\frac37\sqrt{21}} \right)  \right]-h_0
   \end{multline}
%\en
which is very close to Treloar's approximation (\ref{e25a}).

\subsection{Puso's model}
\label{sec:Puso}

Puso \cite[Eq.\@ (1.2.7)]{puso} approximates the inverse Langevin function by\footnote{A simple pole at $x=1$  with residue $- 3/2$ instead of $-1$.}
\be  \label{f22}   \mathscr{L}^{-1}(x) \approx  \frac{3x}{1-x^3}  \en
leading to the response function, see \cite[ Eq.\@  (5.3)]{beatty2008},
\be\label{a28} \betah_{\rm Puso} = \frac{\mu}{1-x^3}, \en
so that $\mu_0\approx 1.01\mu$  with $\im=60$.
The strain energy, see \cite[ Eq.\@  (7.3)]{beatty2008}, is
\be \label{f23} W_{\rm Puso} =  \frac{1}{6} \mu\im \left[\log\left(\frac{1+x+x^2}{(1-x)^2}\right)
   - 2\sqrt{3}\arctan\left(\frac{1+2x}{\sqrt{3}} \right)  \right]-h_0,
   \en
see also \cite[Section 9]{beatty2003}. By expanding Eq.\@  (\ref{a28}), first in partial fractions and then as a Taylor series, we obtain
\be\label{f24}
\mu^{-1} \betah_{\rm Puso} = \frac{1/3}{1-x} +\frac23-\frac13x-\frac13x^2+\cdots  \en
where the first term captures exactly the pole at $x=1$ and the rest is an infinite series.  We may similarly subtract out the pole contribution at $x=1$ from stress response (\ref{eq42}) of the Arruda-Boyce eight-chain model to obtain
\be\label{f25}
\mu^{-1}\betah_{\rm 8ch} = \frac{1/3}{1-x} +\frac23-\frac13x+\frac{4}{15}x^2-\cdots.
\en
The pole term and the first two terms of the series are identical in Eqs.\@ (\ref{f24}) and (\ref{f25}) which goes some way towards explaining the agreement between the two models.

\subsection{Indei et al.\@'s model}
\label{sec:Indei}
Indei et al.\@ \cite[Eq.\@ (13)]{indei} proposed the following model to approximate the inverse Langevin function:
\[
\mathscr{L}^{-1}(x) \approx  3x\left({1+\frac{2A}{3}\frac{x^2}{1-x^2}}\right),
\]
in which $A=1$ gives Cohen's approximation Eq.\@  (\ref{e13}),  $A= 9/10$ gives our new approximation Eq.\@ (\ref{e17}), $A=0$ gives the neo-Hookean model Eq.\@ (\ref{eq14}) and  $A=3/2$ gives Warner's model Eq.\@ (\ref{eq23}).

%\clearpage
\subsection{Models based on the additive removal of the real singularities of $ \mathscr{L}^{-1}(x)$.}
We have considered models of rubber elasticity based on the reduced inverse Langevin function $f(x)$ defined by Eq.\@ (\ref{eq28}) which was obtained by \emph{multiplying} out the real singularities of the inverse Langevin function.  Instead,
we now decompose $ \mathscr{L}^{-1}(x) $ \emph{additively} as
\be
 \mathscr{L}^{-1}(x) = \frac{2x}{1-x^2} +g(x)
 \label{eqt1}
\en
in which the first term consists of the simple poles of $ \mathscr{L}^{-1}(x)$ at $x=\pm1$, each with residue $-1$, and the second term is
\be
g(x) = x-\frac15x^3-\frac{53}{175}x^5+\cdots
\label{eqt3}
\en
where each coefficient in (\ref{eqt3}) is exactly 2 less than the corresponding coefficient in (\ref{eq27}).
The function $g(x)$ has no  singularities at $x=\pm1$ and the series has the same radius of convergence as the series (\ref{eq27}) and (\ref{eq28}).

 Taking only the first term of the series (\ref{eqt3}) gives the approximation
\be\label{eqn10}
 \mathscr{L}^{-1}(x) \approx \frac{2x}{1-x^2}+x
 \en
 which has the correct behaviour as $x\to\pm1$ and   $ \mathscr{L}^{-1}(x) \approx 3x$ as $x\to 0$.  In fact, this approximation is identical to the approximation (\ref{e13}) of Cohen's model.  Formerly, Cohen's approximation (\ref{e13}) was derived in an ad hoc manner by rounding certain coefficients in the [3/2] Pad\'e approximant of the inverse Langevin function but here we see it derived in a more rational manner.  It now becomes clear why Cohen's model has exactly the right singular behaviour as $x\to\pm1$.
 
 Taking the first \emph{two} terms of  (\ref{eqt3}) gives the approximation
  \be\label{eqn11}
 \mathscr{L}^{-1}(x) \approx \frac{2x}{1-x^2}+x -\frac15x^3
 \en
and taking the [1/2] Pad\'e approximant of $g(x)$ gives the  approximation
\be\label{eqn12}
 \mathscr{L}^{-1}(x) \approx \frac{2x}{1-x^2}+  \frac{x}{1+\frac15x^2}.
 \en
However, it turns out that these two further approximations lead to models not much more accurate than Cohen's.

\section{Numerical comparison of the various models} % section 7
\label{sec:numerical}

In Figures \ref{fig:12wwwca} and \ref{fig:13wwwb} we compare the stress response and strain energy, respectively, of those limited-stretch models of rubber elasticity which are not regarded as approximations to the Arruda-Boyce \cite{arruda} eight-chain model or the neo-Hookean model of rubber elasticity, both of which are based on statistical mechanics.  These models are those of Gent \cite{gent}, Beatty \cite{beatty2008}, Van der Waals \cite{beatty2008} and \cite{kilian},  and Warner \cite{warner}, which are discussed in Section  \ref{sec:various}. 

The stress response $\beta$ is depicted in Figure \ref{fig:12wwwca}  for each of the following models:   Neo-Hookean, Eq.\@  (\ref{eq15}); Gent, Eq.\@  (\ref{eq17}); Beatty, Eq.\@  (\ref{eq18}); Van der Waals', Eq.\@  (\ref{eq21});  Warner, Eq.\@  (\ref{eq22}); Arruda-Boyce, Eq.\@  (\ref{eq42}).  For each model we take the number of polymer links in each chain to be $N=20$ for the sake of definiteness.  We employ this value in all our numerical illustrations.  It follows from Eq.\@ (\ref{eq10}) that the maximum value $\im$  which the first principal invariant $I_1$ can take in each Figure is $\im=60$ so that  $3\leq I_1 < \im$.
We see in Figure \ref{fig:12wwwca} that the neo-Hookean stress response is a horizontal line consistent with the fact that it is constant, see Eq.\@  (\ref{eq15}). 

\begin{figure}[H]% Fig. 6.  \beta
\centerline{
\begin{tikzpicture}
\node (0,0) {\includegraphics[scale=0.75]{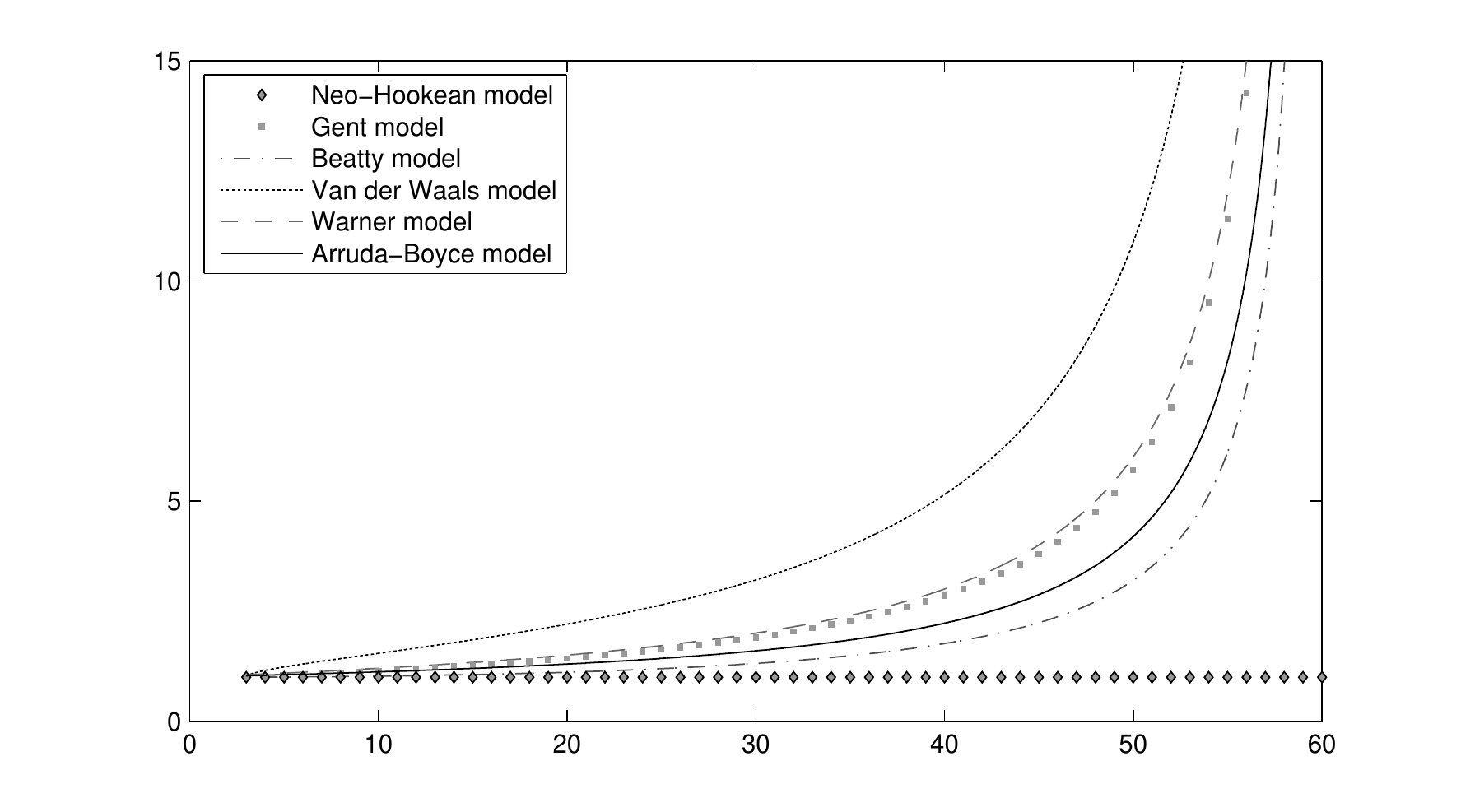}};
\draw  (-5.8,0.1) node [rotate=90] {\fontsize{10}{10} $\beta(I_1)/\mu$};
\draw  (0.3,-3.5) node {\fontsize{10}{10} $I_1$};
\end{tikzpicture}}  
\vspace{-5pt}
\caption{{Stress response.  Graphs of the stress response $\beta$ for each of the models:  Neo-Hookean, Eq.\@  (\ref{eq15}); Gent, Eq.\@  (\ref{eq17}); Beatty, Eq.\@  (\ref{eq18}); Van der Waals, Eq.\@  (\ref{eq21});  Warner, Eq.\@  (\ref{eq22}); Arruda-Boyce eight-chain, Eq.\@  (\ref{eq42}).
 In each case,  $3\leq I_1 < \im$ with $\im=60$.}} 
\label{fig:12wwwca}
\end{figure}

The  strain energy $W$ is depicted in Figure \ref{fig:13wwwb} for the same models of rubber elasticity as are depicted in Figure \ref{fig:12wwwca}.  We see in Figure \ref{fig:13wwwb} that the neo-Hookean strain energy  is a straight line of slope 1/2, consistently with  Eq.\@  (\ref{eq14}).  Beatty's is the only limited-stretch model of rubber elasticity to predict a strain energy less than that of the Arruda-Boyce model.  Warner's strain energy is slightly greater than Gent's, as is clear from Eq.\@   (\ref{eq24}).

\begin{figure}[H]% Fig. 7.  W 
\centerline{
\begin{tikzpicture}
\node (0,0) {\includegraphics[scale=0.75]{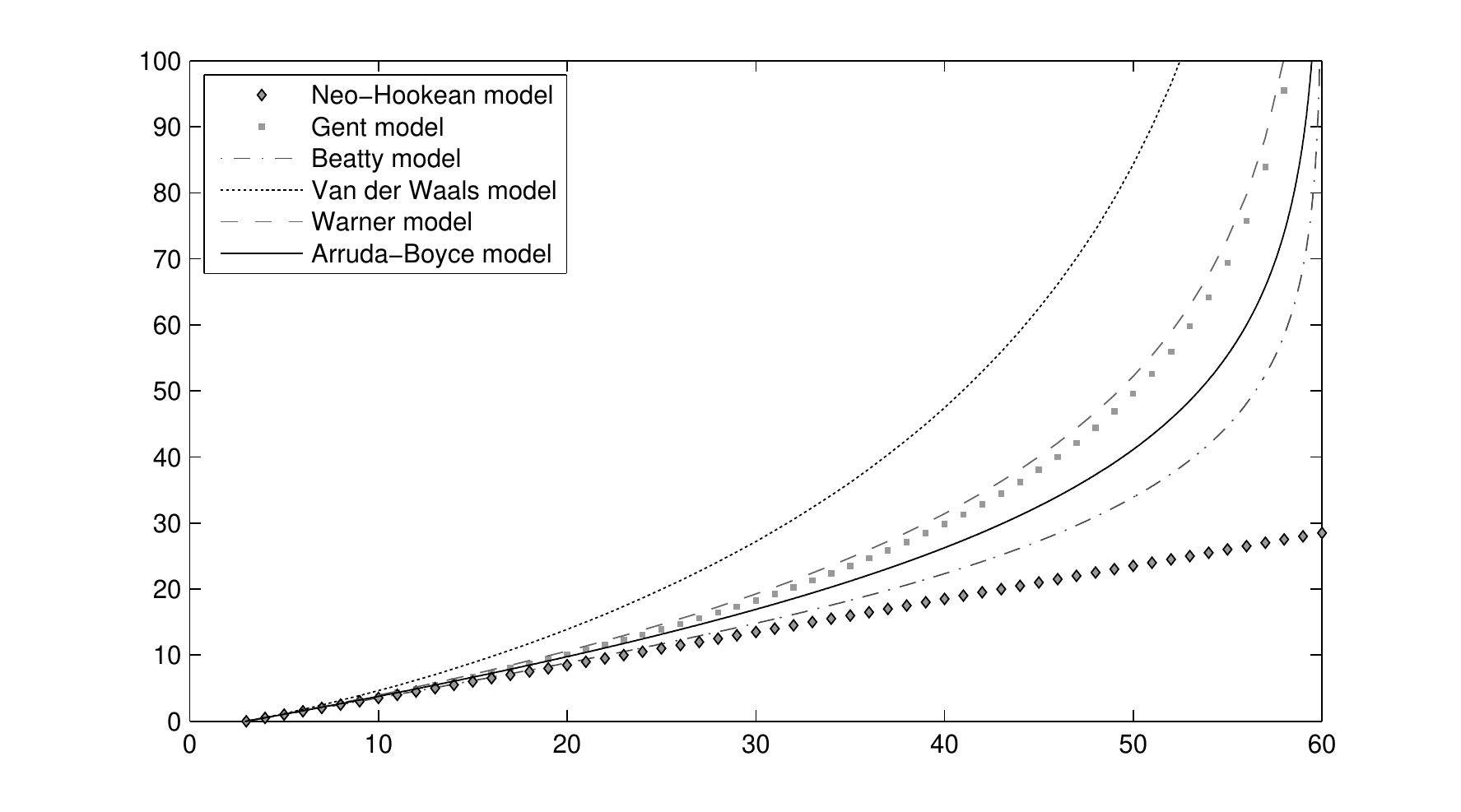}};
\draw  (-5.8,0.1) node [rotate=90] {\fontsize{10}{10} $W(I_1)/\mu$};
\draw  (0.3,-3.5) node {\fontsize{10}{10} $I_1$};
\end{tikzpicture}}  
\vspace{-5pt}
\caption{{Strain energy.  Graphs of the strain energy $W/\mu$ for each of the models:  Neo-Hookean, Eq.\@  (\ref{eq14}); Gent, Eq.\@  (\ref{eq16}); Beatty, Eq.\@  (\ref{eq19});  Van der Waals, Eq.\@  (\ref{eq20}); Warner, Eq.\@  (\ref{eq23}); Arruda-Boyce eight-chain, Eq.\@  (\ref{eq43}).
 In each case,  $3 \leq I_1<\im$ with $\im=60$.}
}
\label{fig:13wwwb}
\end{figure}

For the rest of this section we consider only those models which may be regarded as approximations to the Arruda-Boyce model.

Figure \ref{fig:12wwwc} depicts the stress response in those models of rubber elasticity that may be regarded as approximating the Arruda-Boyce \cite{arruda} eight-chain model; these are the models of Puso \cite{puso}, Cohen \cite{cohen}, our new model Eq.\@  (\ref{e18}), Treloar \cite{treloar1975}, and our modification of Treloar's model  Eq.\@  (\ref{a27}).  Figure \ref{fig:12wwwb}  repeats this for the strain energies in each of the models.  In both figures we see that the models all agree pretty well with each other and with the Arruda-Boyce model.  We shall examine the precise degree of agreement between all these models later in Table \ref{table:2}.

\begin{figure}[H]% Fig. 8.  \beta
\centerline{
\begin{tikzpicture}
\node (0,0) {\includegraphics[scale=0.75]{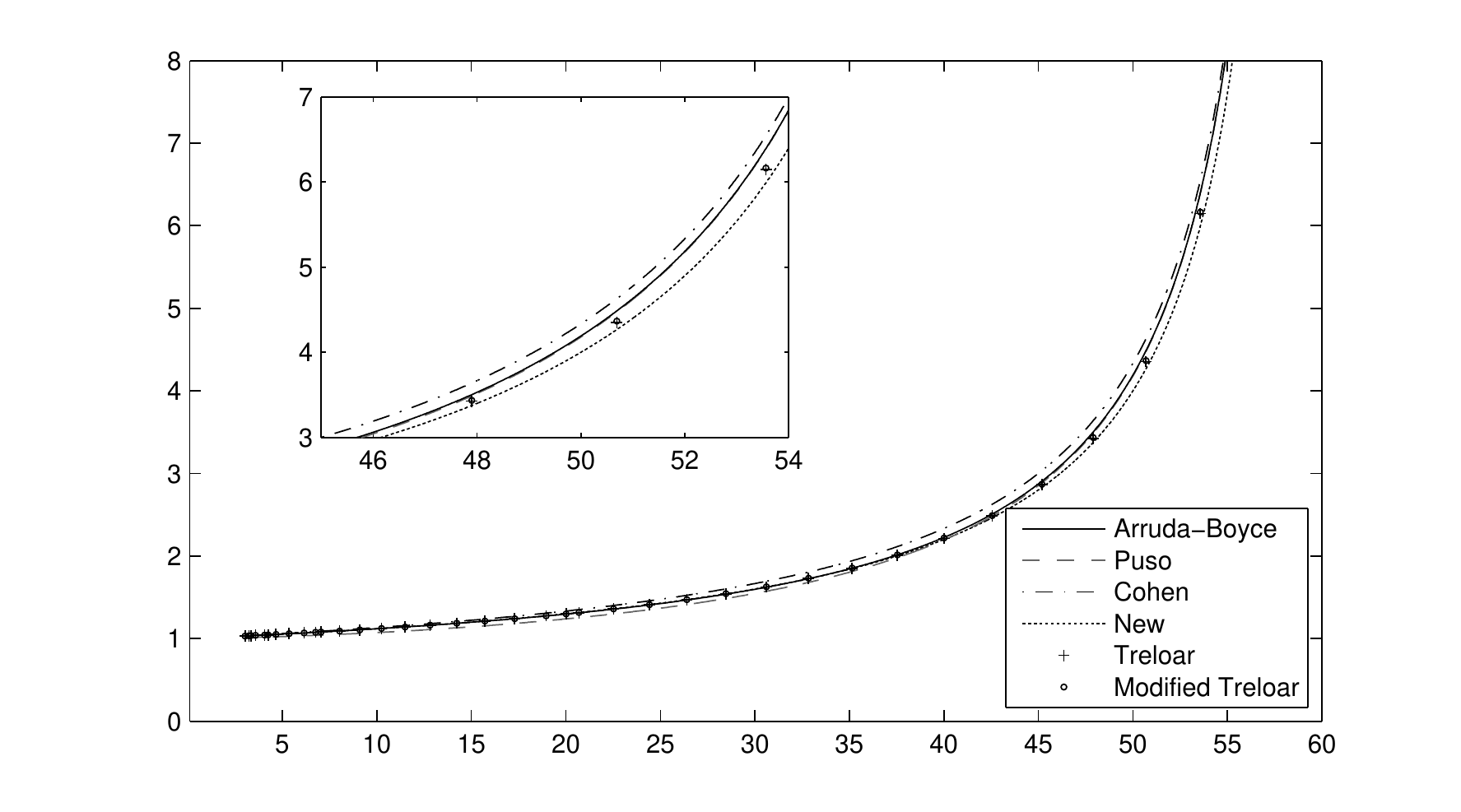}};
\draw  (-5.8,0.1) node [rotate=90] {\fontsize{10}{10}  $\beta(I_1)/\mu$};
\draw  (0.3,-3.5) node {\fontsize{10}{10} $I_1$};
\end{tikzpicture}}  
\vspace{-5pt}
\caption{{Stress response.  Graphs of the stress response $\beta$ for each of the models:  Arruda-Boyce, Eq.\@  (\ref{eq42}); Puso, Eq.\@  (\ref{a28});  Cohen, Eq.\@  (\ref{e14}); New, Eq.\@  (\ref{e18}); Treloar, Eq.\@  (\ref{e24}); modified Treloar, Eq.\@  (\ref{a27}).
 In each case,  $3\leq I_1 < \im$ with $\im=60$. The subfigure shows the small divergence between the   models when $44\leq I_1\leq54.$}}
\label{fig:12wwwc}
\end{figure}

\begin{figure}[h]% Fig. 9.  W 
\centerline{
\begin{tikzpicture}
\node (0,0) {\includegraphics[scale=0.75]{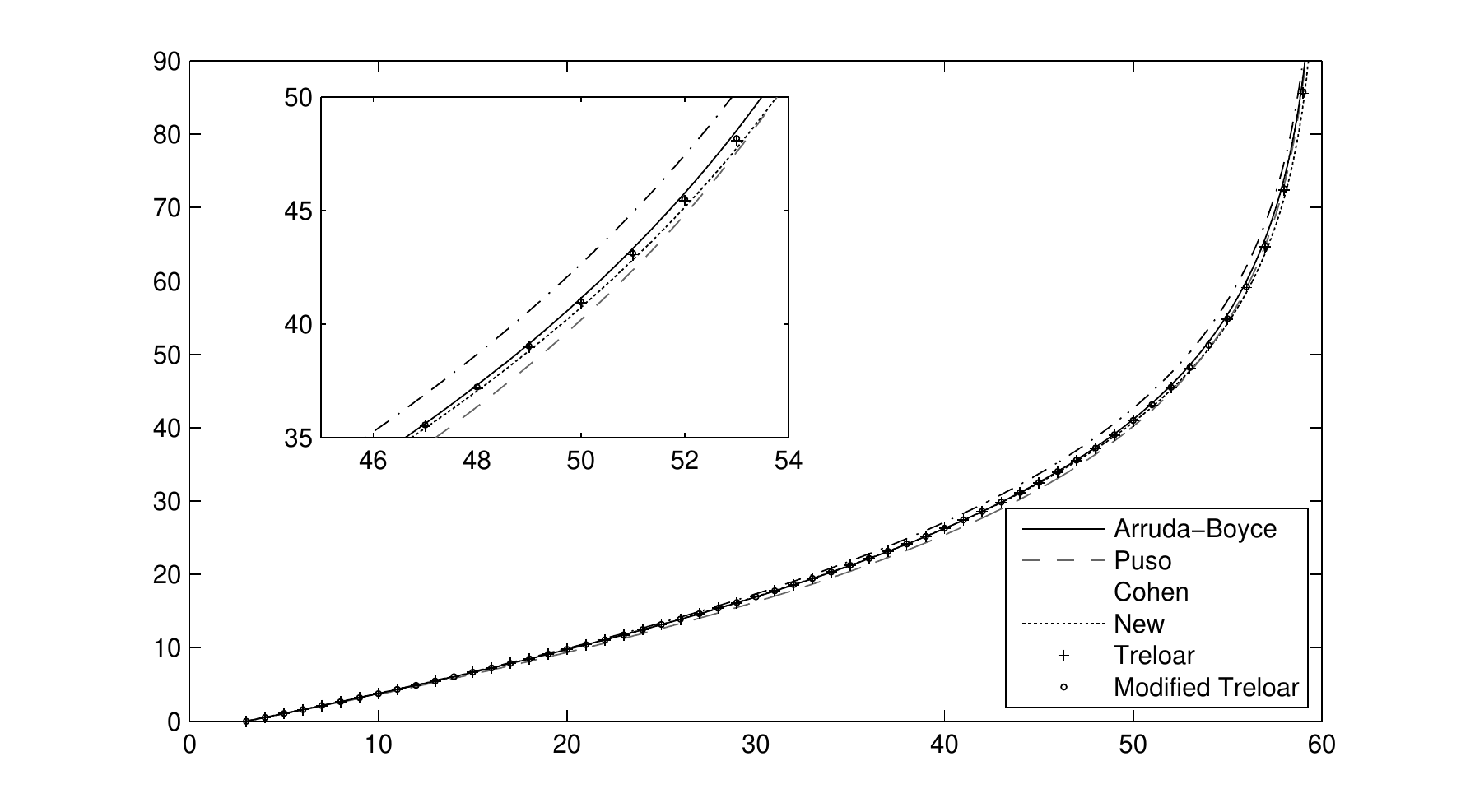}};
\draw  (-5.8,0.1) node [rotate=90] {\fontsize{10}{10} $W(I_1)/\mu$};
\draw  (0.3,-3.5) node {\fontsize{10}{10} $I_1$};
\end{tikzpicture}}  
\vspace{-5pt}
\caption{{Strain energy.  Graphs of the strain energy $W$ for each of the models:   
Arruda-Boyce, Eq.\@  (\ref{eq43});  Puso, Eq.\@  (\ref{f23});   Cohen, Eq.\@  (\ref{e16}); New, Eq.\@  (\ref{e19}); Treloar, Eq.\@  (\ref{e25a}); modified Treloar, Eq.\@  (\ref{e27}).
 In each case,  $3 \leq I_1<\im$ with $\im=60$. The subfigure shows the small divergence between the   models when $44\leq I_1\leq54.$}
}
\label{fig:12wwwb}
\end{figure}

Figure \ref{fig:12ww} depicts the uniaxial tension $T_{11}^{\rm uni}$ calculated using the Arruda-Boyce stress response (\ref{eq42}) together with the uniaxial tensions calculated for each of the models of Puso \cite{puso}, Cohen \cite{cohen}, our new model Eq.\@  (\ref{e18}), Treloar \cite{treloar1975}, and our modification of Treloar's model  Eq.\@  (\ref{a27}).  It can be seen that they agree very much with each other and this is explored further in Table  \ref{table:2}.

Figures \ref{fig:12wwa}, \ref{fig:12wwb} and \ref{fig:12wwc} compare biaxial tension, pure shear and simple shear, respectively, for the same models.   All models are in close agreement, see also Table   \ref{table:2}.

\begin{figure}[H]% Fig. 10.  T^{uni}_{11}
\centerline{
\begin{tikzpicture}
\node (0,0) {\includegraphics[scale=0.75]{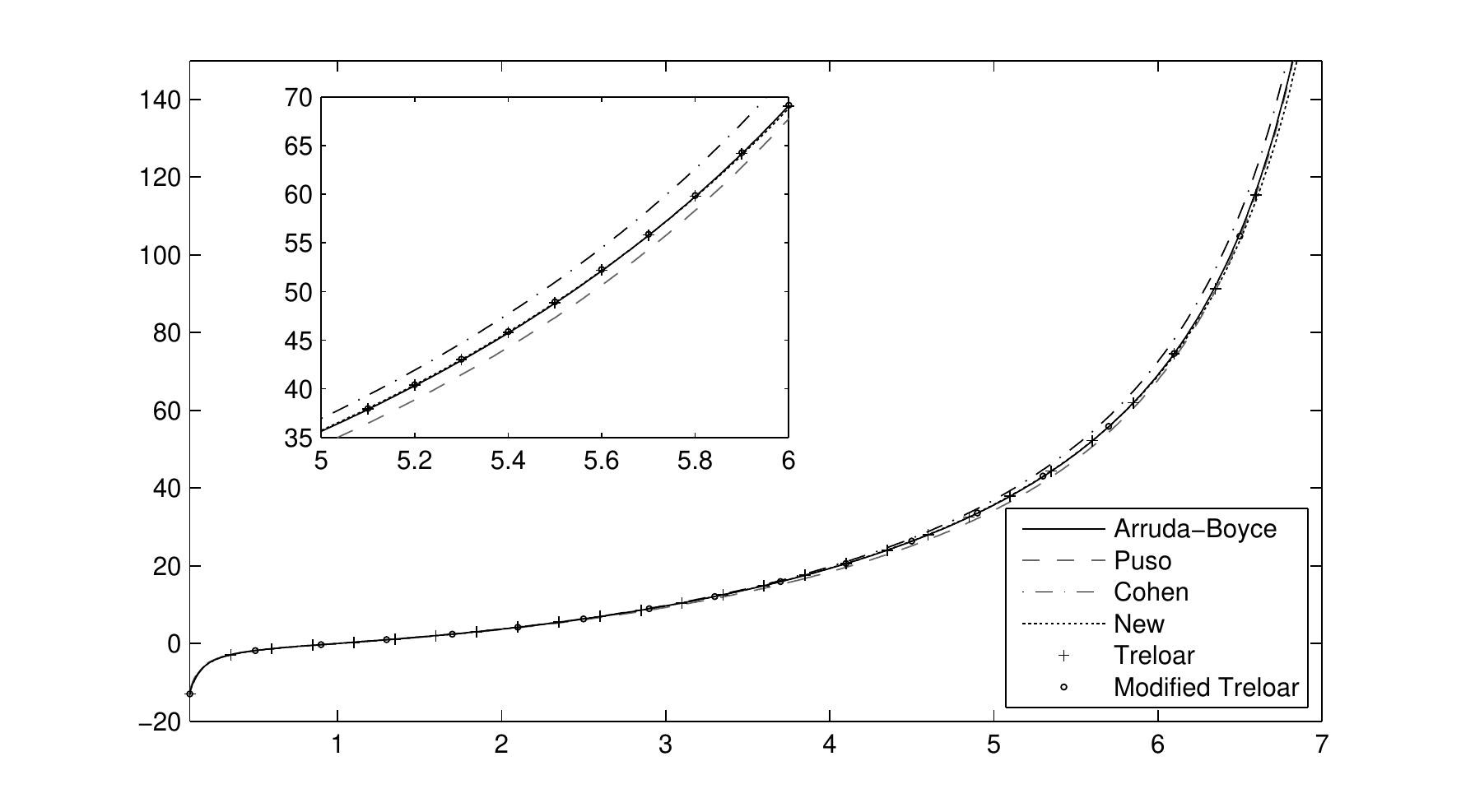}};
\draw  (-5.8,0.1) node [rotate=90] {\fontsize{10}{10} $T_{11}^{\rm uni}(\lambda)/\mu$};
\draw  (0.3,-3.5) node {\fontsize{10}{10} $\lambda$};
\end{tikzpicture}}  
\vspace{-5pt}
\caption{{Uniaxial tension.  Graphs of the uniaxial tension $T_{11}^{\rm uni}$, see Eq.\@  (\ref{eq5}), with the appropriate stress response $\beta$  for the model:  Arruda-Boyce, Eq.\@  (\ref{eq42}); Puso, Eq.\@  (\ref{a28}); Cohen, Eq.\@  (\ref{e14}); New, Eq.\@  (\ref{e18}); Treloar, Eq.\@  (\ref{e24}); modified Treloar,  Eq.\@  (\ref{a27}).
In each case  $0.15\leq\lambda\leq7$ and $\im=60$. The subfigure shows the small divergence between the   models when $5\leq\lambda\leq6$}.}
\label{fig:12ww}
\end{figure}

\begin{figure}[H]% Fig. 11.  T^{bi}_{11}
\centerline{
\begin{tikzpicture}
\node (0,0) {\includegraphics[scale=0.75]{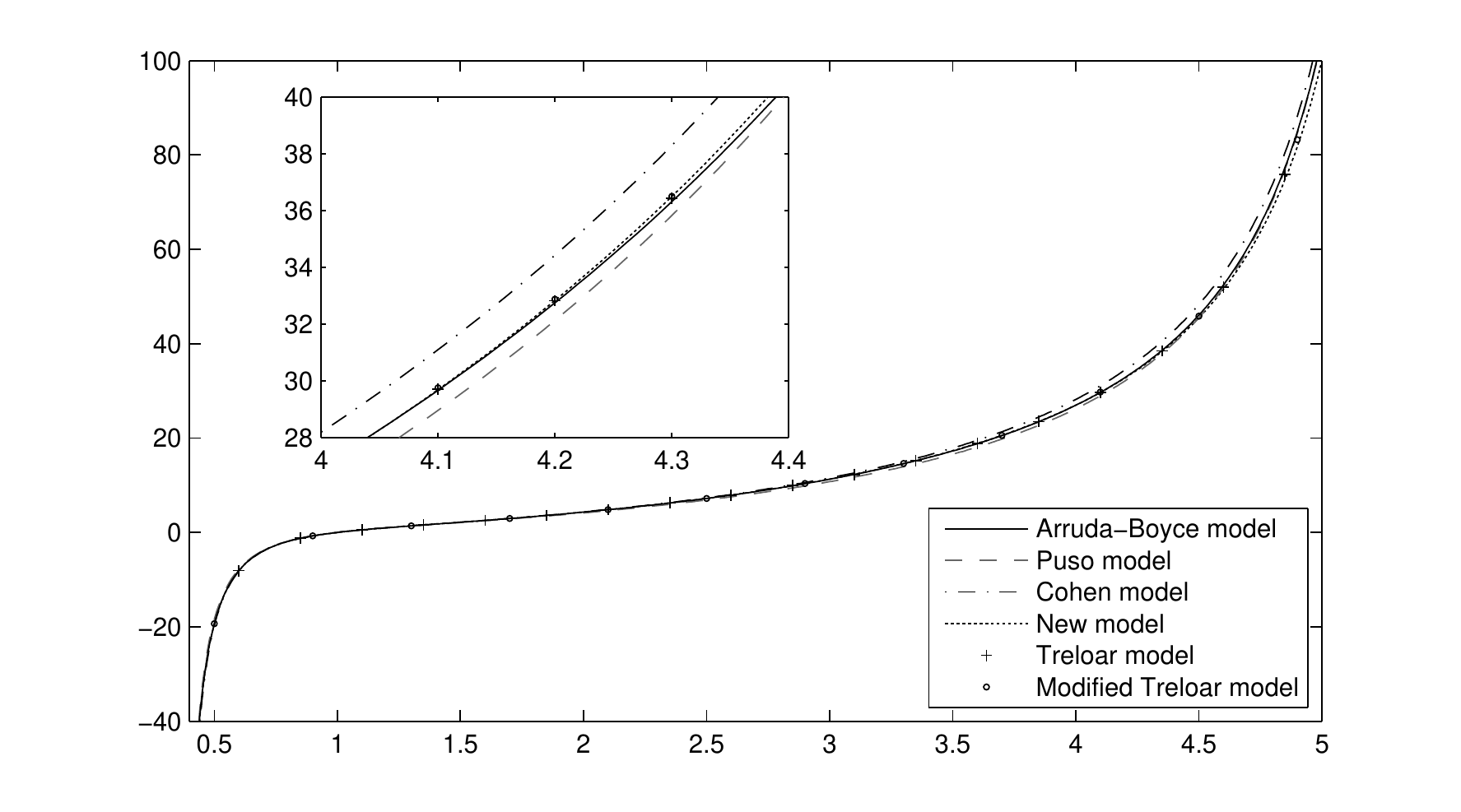}};
\draw  (-5.8,0.1) node [rotate=90] {\fontsize{10}{10} $T_{11}^{\rm bi}(\lambda)/\mu$};
\draw  (0.3,-3.5) node {\fontsize{10}{10} $\lambda$};
\end{tikzpicture}}  
\vspace{-5pt}
\caption{{Biaxial tension.   Graphs of the biaxial tension $T_{11}^{\rm bi}=T_{22}^{\rm bi}$, see Eq.\@  (\ref{eq6}),  with the appropriate stress response $\beta$  for the  model:  Arruda-Boyce, Eq.\@  (\ref{eq42}); Puso, Eq.\@  (\ref{a28}); Cohen, Eq.\@  (\ref{e14}); New, Eq.\@  (\ref{e18}); Treloar, Eq.\@  (\ref{e24}); modified Treloar, Eq.\@  (\ref{a27}).
In each case,  $0.4\leq\lambda\leq5$ and $\im=60$.   The subfigure shows the small divergence between the   models when $4\leq\lambda\leq4.4$.}
}
\label{fig:12wwa}
\end{figure}

\begin{figure}[H]% Fig. 12.  T^{ps}_{11}
\centerline{
\begin{tikzpicture}
\node (0,0) {\includegraphics[scale=0.75]{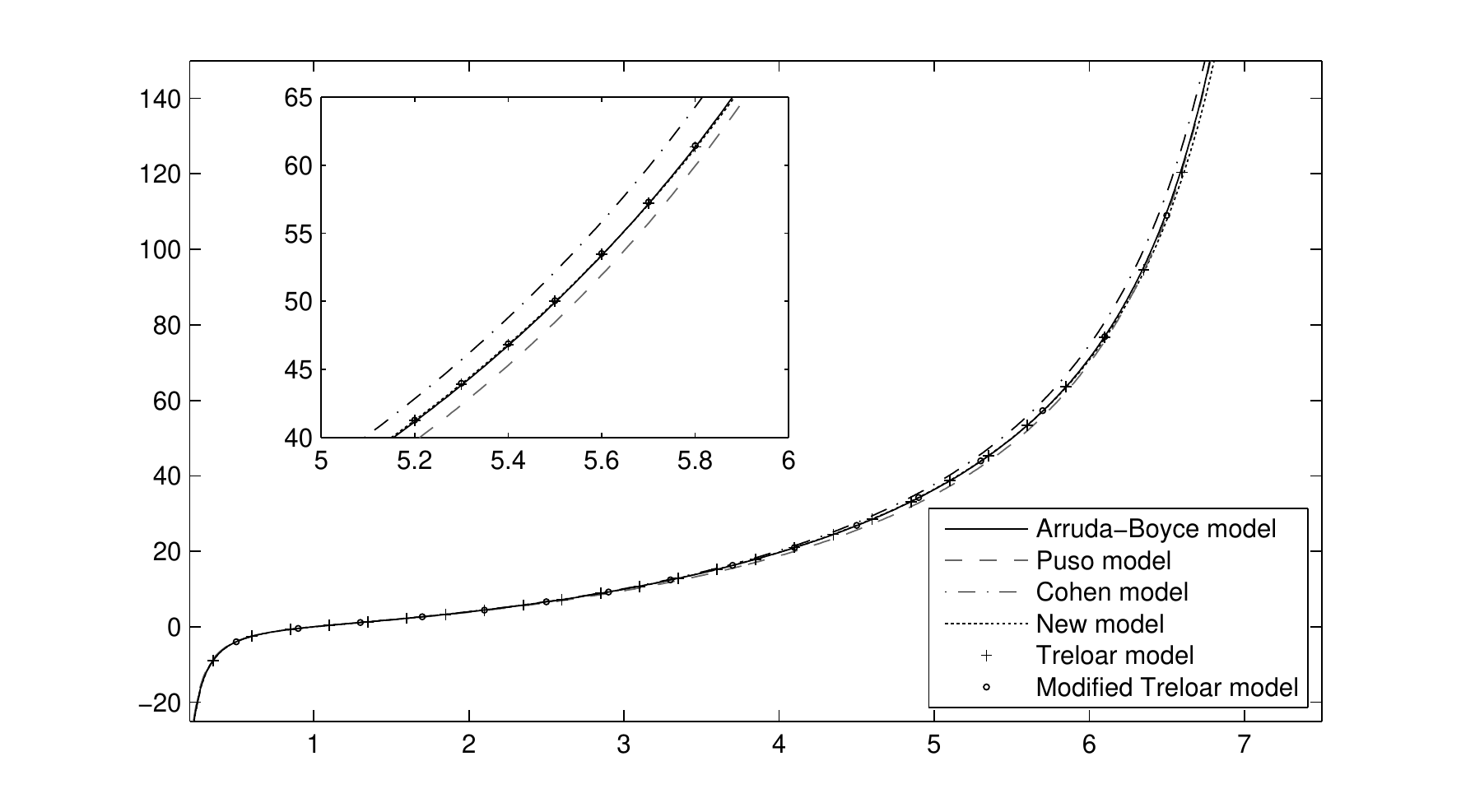}};
\draw  (-5.8,0.1) node [rotate=90] {\fontsize{10}{10} $T_{11}^{\rm ps}(\lambda)/\mu$};
\draw  (0.3,-3.5) node {\fontsize{10}{10} $\lambda$};
\end{tikzpicture}}  
\vspace{-5pt}
\caption{{Pure shear.   Graphs of the normal stress  $T_{11}^{\rm ps}$ in pure shear, see Eq.\@  (\ref{eq7}),  with the stress response $\beta$ chosen to be appropriate for the different models:  Arruda-Boyce, Eq.\@  (\ref{eq42}); Puso, Eq.\@  (\ref{a28}); Cohen, Eq.\@  (\ref{e14}); New, Eq.\@  (\ref{e18}); Treloar, Eq.\@  (\ref{e24}); modified Treloar, Eq.\@  (\ref{a27}).  In each case,  $0.2\leq\lambda\leq7$ and $\im=60$.
The subfigure shows the small divergence between the   models when $5\leq\lambda\leq6$}.
}
\label{fig:12wwb}
\end{figure}

\begin{figure}[H]% Fig. 13.  T^{ss}_{11}
\centerline{
\begin{tikzpicture}
\node (0,0) {\includegraphics[scale=0.75]{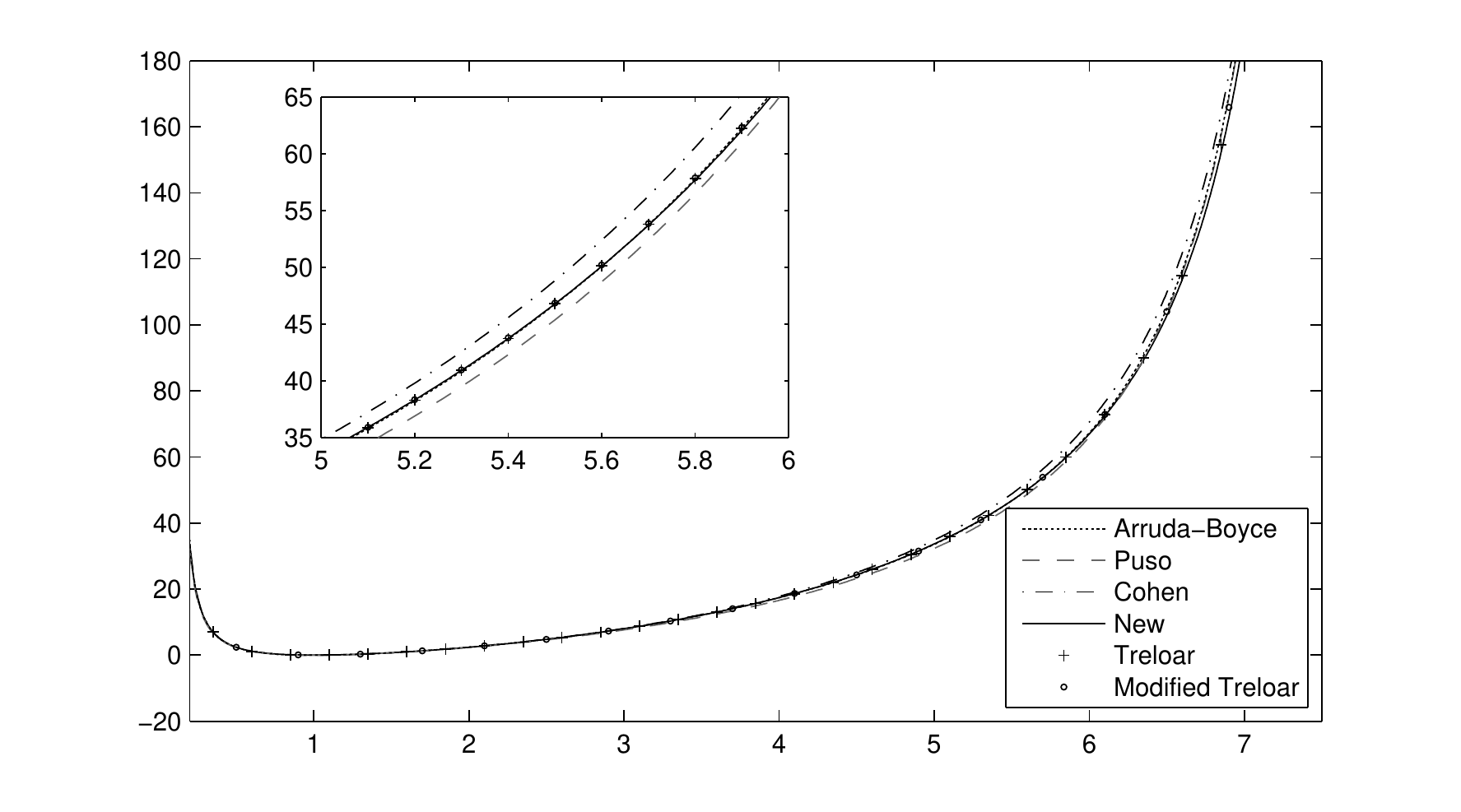}};
\draw  (-5.8,0.1) node [rotate=90] {\fontsize{10}{10} $T_{11}^{\rm ss}(\lambda)/\mu$};
\draw  (0.3,-3.5) node {\fontsize{10}{10} $\lambda$};
\end{tikzpicture}}  
\vspace{-5pt}
\caption{{Simple shear.   Graphs of the normal stress $T_{11}^{\rm ss}$   in simple shear, see Eq.\@  (\ref{eq8}),  with the appropriate  stress response $\beta$  for the  model:  Arruda-Boyce, Eq.\@ (\ref{eq42}); Puso, Eq.\@  (\ref{a28}); Cohen, Eq.\@  (\ref{e14}); New, Eq.\@  (\ref{e18}); Treloar, Eq.\@  (\ref{e24}); modified Treloar, Eq.\@  (\ref{a27}).
In each case,  $0.2\leq\lambda\leq7$ and $\im=60$.   The subfigure shows the small divergence between the   models when $5\leq\lambda\leq6$}.
}
\label{fig:12wwc}
\end{figure}

\begin{table}[H]% Table 1.  
\centering
%\begin{tabular}{c c c c c c c}
\begin{tabular}{l l l l l l l}
\hline\hline 
            & Stress              & Strain  &  Uniaxial                                              & Biaxial      & Pure  & Simple   \\ 
Model & response      & energy  &  tension\hspace{1mm}(\ref{eq5})   & tension\hspace{1mm}(\ref{eq6})  
            & shear\hspace{1mm}(\ref{eq7}) &  shear\hspace{1mm}(\ref{eq8})   \\ [0.5ex] 
\hline
% Warner               &  26.30\hfill (\ref{eq22}) & 14.55  &15.81  & 15.29  & 15.29 &  16.11\hspace{3mm}(\ref{eq23}) \\
 Puso                &  2.48\hfill (\ref{a28}) &  3.14\hspace{3mm}(\ref{f23}) & 3.03  &3.06  & 3.04  & 3.04  \\
Cohen                 &  3.01\hfill (\ref{e14}) &  2.32\hspace{3mm}(\ref{e16}) & 2.26  &2.43  & 2.35  & 2.35  \\
New                     &  1.90 \hfill (\ref{e18}) &  0.58\hspace{3mm}(\ref{e19}) & 0.39  &0.45  & 0.43  & 0.43   \\  
Treloar                 &  1.16\hfill (\ref{e24}) & 0.33\hspace{3mm}(\ref{e25a}) & 0.20  &0.24   & 0.23 & 0.23  \\
Modified& 1.09\hfill (\ref{a27}) &  0.31\hspace{3mm}(\ref{e27}) & 0.20  &0.24  & 0.23  & 0.23    \\
\hfill Treloar  & \\[1ex] 
\hline 
\end{tabular}
\caption{Mean percentage errors of the models of Puso, Cohen, the new model, Treloar and the modified Treloar model as compared with the Arruda-Boyce eight-chain model,  for the  stress response $\beta$, the strain energy $W$ and the stress  $T_{11}$ calculated for the four deformations indicated in the last four columns.  The range of $\lambda$ is:  $0.15\leq\lambda\leq7$ for uniaxial tension, pure shear, simple shear and $0.4\leq\lambda\leq5$ for biaxial tension.  The  numbers in parentheses refer to equation numbers in the text.}
\label{table:2}
\end{table}
In the first two columns, Table \ref{table:2} provides  the mean percentage errors, as compared with the Arruda-Boyce model,  for the stress response and strain energy, respectively, for the models of Puso, Eqs.\@ (\ref{a28}) and (\ref{f23}); Cohen,  Eqs.\@ (\ref{e14}) and (\ref{e16}); New,  Eqs.\@ (\ref{e18}) and (\ref{e19}); Treloar,  Eqs.\@ (\ref{e24}) and (\ref{e25a}); modified Treloar,  Eqs.\@ (\ref{a27}) and (\ref{e27}), in the range $3\leq I_1 \leq60$ as illustrated in Figures  \ref{fig:12wwwca} and \ref{fig:12wwwc} for the stress response and Figures \ref{fig:13wwwb} and \ref{fig:12wwwb} for the strain energy. 
 
 The final four columns of Table \ref{table:2} provide the mean percentage errors for the same models  for uniaxial tension $T_{11}^{\rm uni}$ in the range $0.15\leq\lambda\leq7$, biaxial tension $T_{11}^{\rm bi}$ in the range $0.4\leq\lambda\leq5$, pure shear $T_{11}^{\rm ps}$ and simple shear $T_{11}^{\rm ss}$ in the range $0.15\leq\lambda\leq7$, as illustrated in Figures \ref{fig:12ww}, \ref{fig:12wwa}, \ref{fig:12wwb} and \ref{fig:12wwc}, respectively. 
 
 Cohen's model and our new model employ similar approximations to the inverse Langevin function, see Eqs.\@ (\ref{e13}) and (\ref{e17}), respectively.  These are both simple approximations but the difference in their accuracy in approximating the inverse Langevin function of the Arruda-Boyce eight-chain model is striking.  From the second and third rows of Table \ref{table:2} we see that the new model has only about a fifth of the mean percentage error of Cohen's model for the strain energy and the tensions in uniaxial tension, biaxial tension, pure shear and simple shear.  For the stress response, in the first column, the new model has only $2/3$ of the percentage error of Cohen's.  This increased percentage error of the new model is perhaps because we are taking the average over the full range of possible values of $I_1$, namely, $3\leq I_1<60$, whereas for the other entries in Table \ref{table:2} the stretch does not approach its maximum value.  If $I_1$ is restricted so that $3\leq I_1<47.5$, it can be shown that the mean percentage error for Cohen's model is 3.24\% whereas that for the new model is only 0.56\%.  It follows that Cohen's model is a better approximation than the new model close to the singularity of the inverse Langevin function.  The reason for this is clear.  Cohen's model captures the position and the nature of the singularities of the inverse Langevin function exactly; both have simple poles at $x=\pm1$, each with residue $-1$.  The new model also has simple poles at $x=\pm1$ but the residues are now $-9/10$.  This small discrepancy close to the singularities will not have much effect in practice.

\section{Conclusions}
\label{sec:conclusion}

In this paper we have compared various limited-stretch models of rubber elasticity from continuum mechanics and  from statistical mechanics, that are dependent on only  the first invariant $I_1$ of the left Cauchy-Green strain tensor.
Exceptionally, the James and Guth \cite{james} three-chain model depends separately on each of the principal stretches.   All the models discussed have two material constants, a shear modulus $\mu$ and the maximum value $\im$ of the first invariant $I_1$.  It has been well documented in the literature that with only these two material constants good agreement can be obtained between theory and experiment.  For example, Boyce  \cite{boyce1996} directly compares  the Gent and Arruda-Boyce eight-chain models, concluding that they agree well with experiment and with each other.

We have chosen to compare all the limited-stretch models presented here with the Arruda-Boyce \cite{arruda} eight-chain model.  This model has been used as our reference because it compares well with experiment \cite{arruda,boyce1996,boyce} and because Beatty \cite{beatty2003} has shown  that the model is generally valid and not at all dependent on the eight-chain structure. 
The models of Treloar \cite{treloar1975}, Cohen \cite{cohen} and Puso \cite{puso} compare  favourably with the eight-chain model. Directly comparing our new model with Cohen's \cite{cohen} and Puso's \cite{puso} models shows that the new model has the smallest mean  percentage deviation from the Arruda-Boyce model for the stress response, the strain energy and for  uniaxial tension, biaxial tension, pure shear and simple shear. 

From Figure \ref{fig:12wwwc} it is seen for the stress response our new model provides the most accurate representation to the inverse Langevin function over the range $3\leq I_1\leq40$, taking $\im=60$.   In this range our new model has mean percentage error $0.23\%$ . For the range $40<I_1<60$ the accuracy of our new model decreases and  the Puso model gives the most accurate fit in this range,  with mean percentage error $0.61\%$ . The loss of accuracy in our model close to the simple pole at $I_1=\im=60$ is because our  approximation does not have the correct residue at this pole;   the models of Cohen and Puso capture this feature exactly. However, Figures \ref{fig:12ww} -- \ref{fig:12wwc} illustrate the small deviation of our new model from the Arruda-Boyce model over most of the $I_1$ range for  uniaxial tension, biaxial tension, pure shear and simple shear, respectively.   Also, Figures \ref{fig:12ww} -- \ref{fig:12wwc} show that  Cohen's model is closer to the Arruda-Boyce model than is Puso's.   All this is evident from Table \ref{table:2}.

From Table \ref{table:2} we see that Puso's and Cohen's models are the least accurate but that our new model, which is as simple as Cohen's, has mean percentage error only about a fifth of  Cohen's.  Treloar's  \cite{treloar1975} model has percentage error about a half of ours but is a much more complicated model.  Our modified Treloar model is only slightly more accurate  than Treloar's for the stress response and strain energy but otherwise shares the same accuracy.

From Eqs.\@ (\ref{e19}) and (\ref{e16b})  and the small percentage deviations of our new model from the Arruda-Boyce model we may conclude that the Arruda-Boyce model is effectively a linear combination of the Gent model and the new model.  This is a simple structure and shows why Boyce \cite{boyce1996} observed such close agreement between the Gent and Arruda-Boyce models.

In this paper we have presented only isotropic versions of the limited-stretch models.  It is possible to introduce anisotropy by rewriting Eq.\@ (\ref{eq12})$_1$ in terms of anisotropic invariants,  see, for example, Rickaby and Scott \cite{rickaby1,rickaby6}  for the cases of transverse isotropy and orthotropy, respectively.

%\paragraph{Acknowledgements}  We would like to thank a reviewer for new references and constructive comments.

\FloatBarrier

\end{document}